\documentclass[preprint,review,12pt,sort&compress]{elsarticle}
\usepackage[margin=1 in]{geometry}
\usepackage{multirow}
\usepackage[usenames, dvipsnames]{color}
\definecolor{UW}{RGB}{64, 38, 96}

\usepackage{amssymb}

\usepackage{float}

\journal{Composites Part A: Applied Science and Manufacturing}
\begin{document}
\begin{titlepage}

\clearpage\thispagestyle{empty}



\noindent

\hrulefill

\begin{figure}[h!]

\centering

\includegraphics[width=1.5 in]{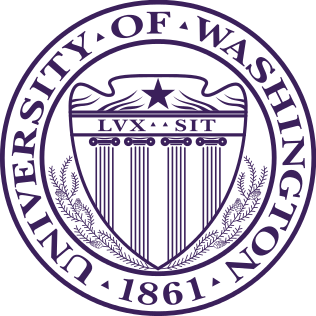}

\end{figure}

\begin{center}

{\color{UW}{

{\bf A\&A Program in Structures} \\ [0.1in]

William E. Boeing Department of Aeronautics and Astronautics \\ [0.1in]

University of Washington \\ [0.1in]

Seattle, Washington 98195, USA

}

}

\end{center} 

\hrulefill \\ \vskip 2mm

\vskip 0.5in

\begin{center}

{\large {\bf A Study on the Multi-axial Fatigue Failure Behavior of Notched Composite Laminates}}\\[0.5in]

{\large {\sc Yao Qiao, Antonio Alessandro Deleo, Marco Salviato}}\\[0.75in]

{\sf \bf INTERNAL REPORT No. 19-07/03E}\\[0.75in]

\end{center}

\noindent {\footnotesize {{\em Submitted to Composites Part A: Applied Science and Manufacturing \hfill July 2019} }}

\end{titlepage}

\newpage

\begin{frontmatter}


\cortext[cor1]{Corresponding Author, \ead{salviato@aa.washington.edu}}

\title{A Study on the Multi-axial Fatigue Failure Behavior of Notched Composite Laminates}


\author[address]{Yao Qiao}
\author[address]{Antonio Alessandro Deleo}
\author[address]{Marco Salviato\corref{cor1}}

\address[address]{William E. Boeing Department of Aeronautics and Astronautics, University of Washington, Seattle, Washington 98195, USA}

\begin{abstract}
\linespread{1}\selectfont

Composite structures must endure a great variety of multi-axial stress states during their lifespan while guaranteeing their structural integrity and functional performance. Understanding the fatigue behavior of these materials, especially in the presence of notches that are ubiquitous in structural design, lies at the hearth of this study which presents a comprehensive investigation of the fracturing behavior of notched quasi-isotropic [+45/90/$-$45/0]$_{s}$ and cross-ply [0/90]$_{2s}$ laminates under multi-axial quasi-static and fatigue loading. 

The investigation of the S-N curves and stiffness degradation, and the analysis of the damage mechanisms via micro-computed tomography clarified the effects of the multi-axiality ratio and the notch configuration. Furthermore, it allowed to conclude that damage progression under fatigue loading can be substantially different compared to the quasi-static case. 

Future efforts in the formulation of efficient fatigue models will need to account for the transition in damaging behavior in the context of the type of applied load, the evolution of the local multi-axiality ratio, the structure size and geometry, and stacking sequence. By providing important data for model calibration and validation, this study represents a first step towards this important goal.

\end{abstract}


\end{frontmatter}


\section{Introduction}

The last few decades have seen a tremendous increase in the use of composite materials in aerospace \cite{rana,baker}, automotive\cite{Ahmed,kaiser}, and civil \cite{Kar03, Brandt08, Ceccato17, Carloni16, Feo13} applications as well as wind energy production \cite{sorensen,chortis}. While, fostered by the need for lightweight and durable structures, the market of composites will continuously expand in future years, the widespread use of these materials is exposing the need for safe and reliable design rules, especially in the context of the complex multiaxial cyclic stress states during service.

Although far less attention has been devoted to fatigue compared to quasi-static loading, the characterization and modeling of the behavior of composite materials under uniaxial and multiaxial loading have been the subjects of countless advances since the pioneering works of Owen and co-authors \cite{owen1, owen2, owen3, owen4, owen5, owen6}. Still today, these are considered among the most comprehensive experimental investigations of the multiaxial fatigue behavior of polymer composites. A formidable set of data was also provided by Fujii \textit{et al}. \cite{Fuji1, Fuji2, Fuji3, Fuji4}, who investigated the fatigue behavior of woven Glass Fiber Reinforced Polymers (GFRPs) under a variety of multiaxial stress states and even in the presence of stress concentrations induced by notches \cite{Fuji5, Fuji6, Fuji7}. 

A very insightful discussion of fatigue damage in composites under on-axis tension was provided by Talreja in \cite{talreja1,talreja2} who identified three distinct mechanisms: (i) fiber failure at high applied strains with nonprogressive failure, (ii) matrix fatigue cracking and progression by failing fibers or by debonding, and (iii) matrix cracking confined by fibers preventing its propagation. 

The damage mechanisms under multiaxial fatigue were investigated, among others, by Wang \textit{et al}. \cite{wang1, wang2} who conducted cryogenic tests in tension/torsion on tubular specimens. They found that the dominating damage mechanisms included matrix cracking, fiber/matrix debonding, microbuckling of fiber bundles, and delamination. The relative contribution of each mechanism was found to be a function of the multiaxiality ratio. 

By testing cruciform GFRP specimens, Smith and Pascoe \cite{pascoe} identified rectilinear cracking, shear degradation of the fiber/matrix interface, and delamination as the main fatigue mechanisms.
Similar results were obtained on carbon/epoxy composites by Chen and Matthews \cite{matthews}. 

A step towards the quantification of microscale fatigue damage is represented by the work of Adden and Horst \cite{horst} who characterized the crack density evolution in tension/torsion loading in NCF glass/epoxy composites.
In this context, a significant contribution was provided by Quaresimin and co-workers who investigated the damage evolution \textit{in-situ} in various loading configurations and geometries \cite{quaresimin1,quaresimin2,quaresimin3,quaresimin4,quaresimin5,quaresimin6}.     

On the modeling side, an early attempt of capturing the fatigue behavior of composites is due to Sims and Brogdon \cite{SimBro77} who tried to extend the Tsai-Hill \cite{tsaihill} static failure criterion to fatigue by substituting the strength parameters with suitable S-N curves. 

A similar approach was pursued by Francis \textit{et al} \cite{francis} while Fujii and Lin \cite{Fuji4} investigated the extension of the Tsai-Wu criterion \cite{tsaiwu} for the description of tension-torsion fatigue of glass/polyester tubes under different multiaxiality ratios. On similar grounds, several other authors proposed alternative extensions (see e.g. \cite{wafa, kawai, philippidis}). 

An excellent discussion on polynomial criteria for multiaxial fatigue was provided by Quaresimin \textit{et al}. \cite{quaresimin7} who verified their predictive capability through a comprehensive re-analysis of a large bulk of experimental data. They concluded that the accuracy of some multiaxial criteria was fair although, in few cases, largely unsafe predictions were obtained undermining the general validity of the models. They also stressed the importance of an insightful understanding of the local fatigue damage mechanisms and their incorporation in micromechanical models as an answer to the foregoing limitations. 

In this context, a notable example is the Synergistic Damage Mechanics (SDM) approach proposed by Singh and Talreja \cite{talreja3} which combines micro-damage mechanics and continuum damage mechanics to predict the stiffness degradation due to presence of transverse cracks. Thanks to the microscale description of damage, the model can be used for various stacking sequences and loading conditions. In this direction is the outstanding work of McCartney and co-workers \cite{mccartney1,mccartney2,mccartney3,mccartney4,mccartney5,mccartney6} and Lundmark and Varna \cite{lundmark1,lundmark2} on the modeling of matrix microcracking in transverse plies.

Later, Carraro \textit{et al}. \cite{carraro1,carraro2} proposed an analytical model that can capture the microcracking in multidirectional laminates with any stacking sequence. Thanks to analytical framework, the model is extremely robust and inexpensive from the computational point of view while still extremely accurate. 

The foregoing results show the significant progress in the understanding of the fatigue behavior of composites. However, while the uniaxial and multiaxial fatigue response in smooth specimens has been the subject of extensive research, the study of notched structures is yet elusive. 

As a first step towards addressing this important problem, this work presents an experimental investigation of the fracturing behavior of notched quasi-isotropic [+45/90/$-$45/0]$_{s}$ and cross-ply [0/90]$_{2s}$ laminates under multiaxial quasi-static and fatigue loading. To have full control of both the multiaxiality ratio and the notch configuration relative to the fiber orientation and specimen geometry, the tests were conducted leveraging an Arcan rig which enables the application of combinations of nominal normal and shear stresses. Thanks to the synergistic use of stiffness degradation data, Digital Image Correlation (DIC), and X-ray Micro-computed tomography the study sheds more light on the effect of the multiaxial stress state, the layup and notch configuration on the fatigue behavior. 

Before moving to the next sections, it is worth pointing out that all the results reported in this work are presented in terms of nominal stresses. Accordingly, the multiaxiality ratio refers to the ``global" multiaxial state of stress rather than the ``local", which may differ from lamina to lamina. Although it is well known that the damage mechanisms driving the fatigue behavior in composites are controlled by the \textit{in-situ} stress field, the choice of using the nominal stress was made to provide objective results for the calibration and validation of multiaxial fatigue models. In fact, in the presence of a stress raiser, the local stress state is in continuous evolution due to the stress/strain re-distribution associated to the progressive damage occurring in the Fracture Process Zone (FPZ). The calculation of such a stress state depends on the ability of the damage model to capture the main damage mechanisms and their evolution.     

\section{Materials and Methods} 
\subsection{Specimen Preparation}
The unidirectional system used for all the tests was a Glass Fiber Reinforced Polymer (GFRP) by Mitsubishi Composites \cite{Rock}.

The investigated stacking sequences included quasi-isotropic [+45/90/$-$45/0]$_{s}$ and cross-ply [0/90]$_{2s}$ layups. To manufacture the laminates, the prepreg was hand laid-up and then vacuum-bagged by using a Vacmobile mobile vacuum system \cite{pump}. A Despatch LAC1-38A programmable oven was used to cure the panels by ramping up the temperature from room temperature to $275^{\circ}$F in one hour, soaking for one hour, and cooling down to room temperature. After curing, the panels were cut into specimen of about 200 $\times$ 44 mm by using a water-cooled circular saw with a diamond-coated blade. The specimen thickness was about 1.72 mm and the gauge length was about 25 mm.

The effect of an open hole or an intra-laminar central crack on composite materials under multi-axial quasi-static and fatigue loading was studied. To this end, three different geometries, illustrated in Figure \ref{fig:geometry}b-d, were prepared: (1) specimen with a central circular hole of diameter $a_0 = 10$ mm, (2) specimen with a central crack of 10 mm, and (3) specimen with a central crack of 18 mm. The open-hole specimens featured only one stacking sequence of [+45/90/$-$45/0]$_{s}$ and a circular hole drilled by a tungsten carbide drill bit. In contrast, the cracked specimens featured both [0/90]$_{2s}$ and [+45/90/$-$45/0]$_{s}$ as stacking sequence. For the former layup, a crack $a_0=18$ mm was considered while, for the latter $a_0$ was equal to 10 mm. The crack was manufactured by firstly drilling a pre-notch using a 0.4 mm tungsten carbide drill bit and then completing the crack leveraging a 0.4 mm wide diamond-coated saw.

In addition to the forgoing notched specimens, unnotched specimens with different dimensions and layups were tested under quasi-static and fatigue loading conditions. The purpose of these tests was to complete the information on the quasi-static and fatigue behavior provided by the manufacturer. The dimensions of the specimens followed ASTM D3039/D3039M \cite{ASTM} and the details can be found in Table \ref{tab:mechanicalproperties}.

\subsection{Test Method}
\begin{figure} [H]
\center
\includegraphics[scale=0.55]{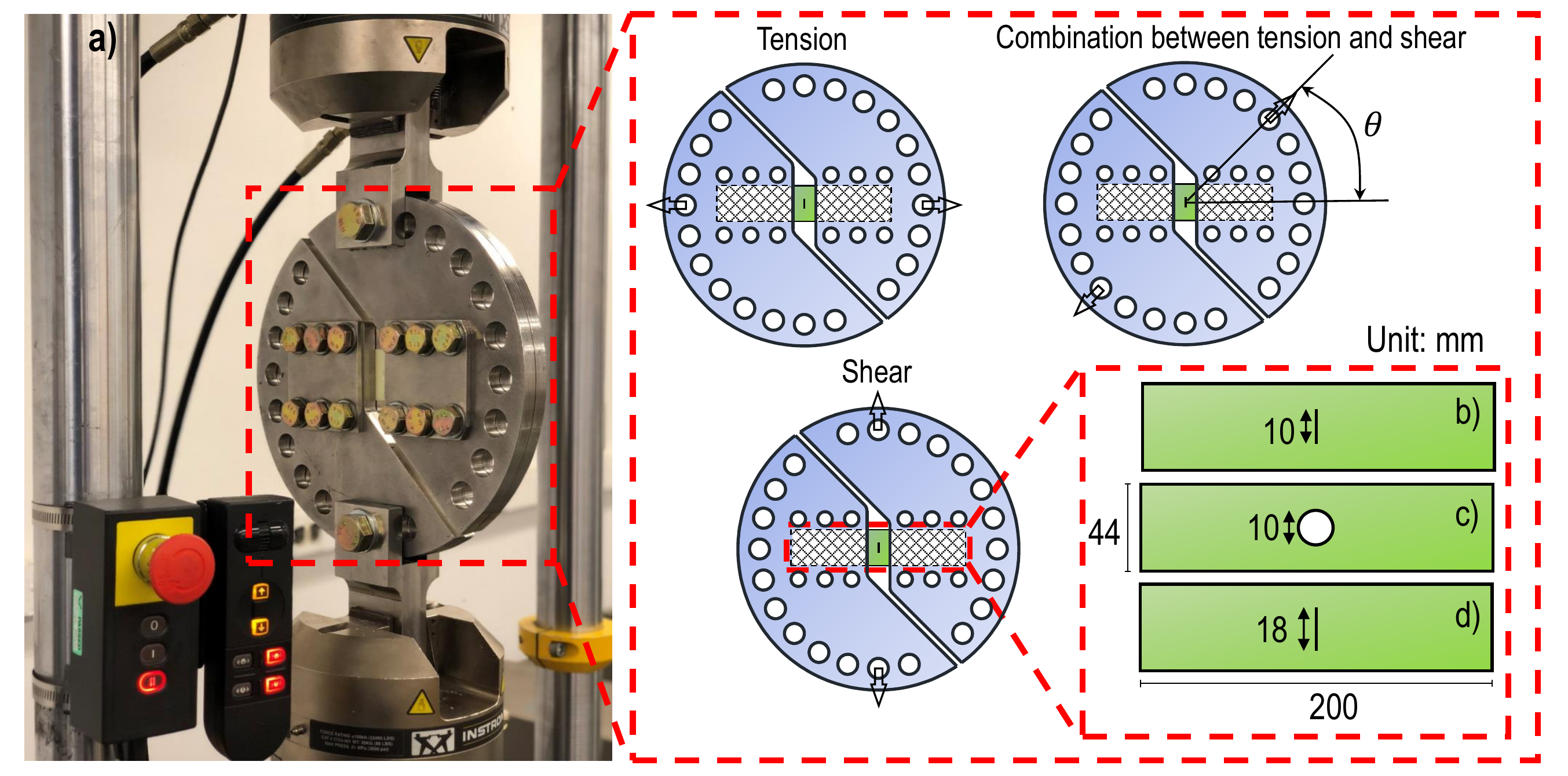}
\caption{(a) Test setup and Arcan rig used for the multi-axial tests. Geometry of notched specimens used for the multi-axial tests: (b) $[+45/90/-45/0]_s$ specimen with a 10 mm central crack, (c) $[+45/90/-45/0]_s$ specimen with a 10 mm hole, and (d) $[0/90]_{2s}$ specimen with a 18 mm central crack. Note that the gauge length is about 25 mm.}
\label{fig:geometry}
\end{figure}

A modified version of Arcan rig, illustrated in Figure \ref{fig:geometry}a, was specifically designed to guarantee infinite life for multi-axial fatigue tests on composite materials. This modified Arcan rig comprises four identical plates (two fronts and two backs) which are used to clamp the specimens by friction. A 17-4 PH stainless steel was used to manufacture these four plates.

The specimens were clamped to the modified Arcan rig by using twelve M14 high-strength bolts, with  each bolt torqued at $130 \mbox{ N m}$. This ensures that enough clamping pressure was applied on the specimen tabs to avoid slipping during the tests. The modified Arcan rig is capable of applying multi-axial loads by varying loading angle $\theta$, the angle between loading direction and the longitudinal direction of the specimen. As illustrated in Figure \ref{fig:geometry}a, tension is applied when $\theta$ equals $0^{\circ}$ while pure shear is applied when $\theta$ equals $90^{\circ}$. A combination of tension and shear is achieved by an intermediate loading angle between $0^{\circ}$ and $90^{\circ}$. To describe the loading configuration, multiaxiality ratio can be defined as $\lambda=$ arctan$(\tau_{N}/\sigma_{N})$ where $\sigma_{N} = P$cos$\theta/[(w-a_0)t]$ is the nominal normal stress and $\tau_{N} = P$sin$\theta/[(w-a_0)t]$ is the nominal shear stress applied to the specimen. In the definitions of stress, $P$ is the instantaneous load, $\theta$ is the loading angle, $w$ is the specimen width, $a_0$ is the crack length or hole diameter and $t$ is the specimen thickness. It is worth mentioning here that the in-plane and out-of-plane bending are negligible in the foregoing tests. This was verified by using multiple strain gauges on the different locations of the specimen or Digital Image Correlation (DIC) on the displacement contours of the specimen under multi-axial loading condition as discussed by Tan \emph{et al.} \cite{Tan}.  

The multi-axial, quasi-static, and fatigue tests were performed in a servo-hydraulic 8801 Instron machine with closed-loop control. The speckled specimens were analyzed by means of an open source Digital Image Correlation system programmed in MATLAB software developed at Georgia Tech \cite{Ncorrverify,Blader}. 

\subsection{Uniaxial tests on notch-free specimens}
\label{sec:tensiletests}
Notch-free specimens were tested under quasi-static and fatigue loading conditions to characterize the uniaxial constitutive behavior and S-N curves. The quasi-static tests were performed under displacement control with a displacement rate of 0.01 mm/s whereas the fatigue tests were under load control with a stress ratio of $R=0.1$ and a low frequency of $f=5$ Hz. It is worth mentioning that both [0/90]$_{2s}$ and [+45/90/$-$45/0]$_{s}$ stacking sequence were investigated. These configurations were the same adopted for the multi-axial tests. The forgoing experimental campaign was performed to provide sufficient information on the uniaxial tensile behavior of the material before investigating the multi-axial behavior.  

\subsection{Multi-axial tests on notched specimens}
\label{sec:multiaxial}
\subsubsection{Multi-axial quasi-static tests}
To study the failure behavior of notched laminates under multi-axial quasi-static loading, five sets of multiaxiality ratios were investigated: $\lambda=$ 0, 0.262, 0.785, 1.309 and 1.571. Such multiaxiality ratios corresponded to $\theta$ = $0^{\circ}$, $15^{\circ}$, $45^{\circ}$, $75^{\circ}$ and $90^{\circ}$. At least three specimens were tested for each configuration. As for the tensile quasi-static tests on unnotched specimens, the load rate for the multi-axial quasi-static tests was 0.01 mm/s.

\subsubsection{Multi-axial fatigue tests}
In the case of multi-axial fatigue tests, the same multiaxiality ratios as the quasi-static tests on notched specimens were investigated. At least three specimens were tested for each multiaxiality ratio in order to produce a statistically significant set of data. To study the fatigue failure behavior of notched laminates under multi-axial fatigue loading, three sets of loading cases were studied: 70\% $P_{max}$, 55\% $P_{max}$ and 40\% $P_{max}$ where $P_{max}$ is the average peak force determined from quasi-static tests for each multiaxiality ratio. A stress ratio of $R=0.1$ and a low frequency of $f=5$ Hz were kept for all the forgoing loading cases.

\subsection{Damage detection and analysis}
A NSI X5000 X-ray micro-tomography scanning system \cite{northstar} was used to observe the sub-critical fatigue damage mechanisms of the specimens with a X-ray tube setting of 90 kV voltage and 220 $\mu$A current. This non-destructive technique was significantly important to guarantee that no additional damage was created during the damage visualization process. In addition, a dye penetrant composed of zinc iodide powder (250 g), isopropyl alcohol (80 ml), Kodak photo-flow solution (1 ml) and distilled water (80 ml) was used in all the scans as a supplement to improve the visualization of the damage mechanisms \cite{dye,dye1}. Prior to the scanning, the specimens were soaked in the dye penetrant mixture for approximately one day. The sub-critical damage in each ply and interface were identified by slicing through the reconstructed 3D images of the specimens leveraging ImageJ software \cite{imagej}.

\section{Experimental Results}
\subsection{Uniaxial tests on notch-free specimens}
\begin{figure} [H]
\center
\includegraphics[scale=0.7]{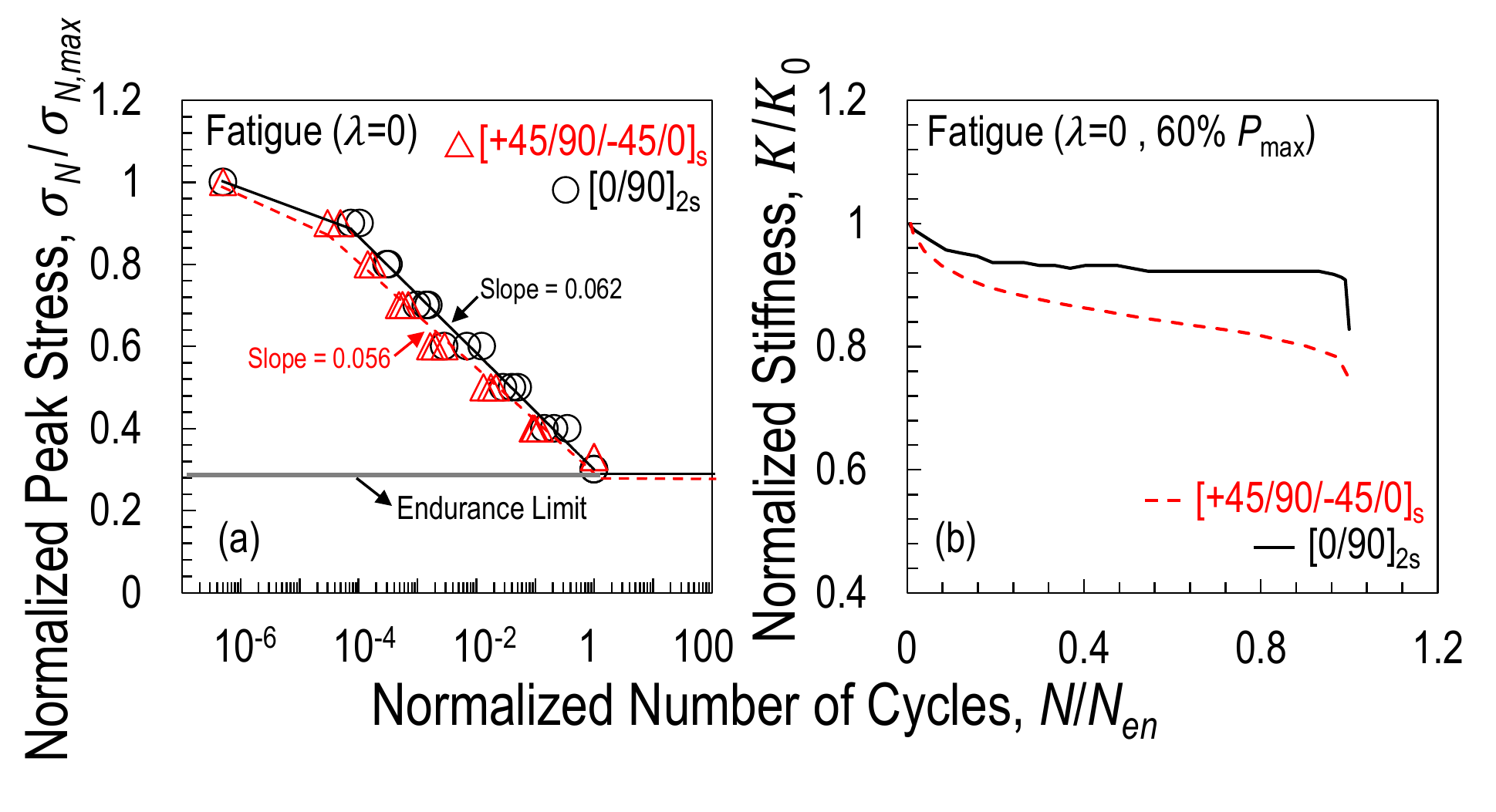}
\caption{(a) Normalized S-N curves for unnotched quasi-isotropic and cross-ply specimens under tensile fatigue loading conditions; (b) Normalized stiffness vs. number of cycles (60\% $P_{max}$). Note that $N_{en}$ equals to 2 million cycles as endurance limit in this work and $\sigma_{N,max}$ represents the quasi-static uniaxial strength of the specimen.}
\label{fig:unnotched}
\end{figure}

The mechanical properties of the unnotched specimens under quasi-static loading conditions are tabulated in Table \ref{tab:mechanicalproperties}. These properties were measured based on the stress and strain curves with the strain obtained from Digital Imaging Correlation (DIC) to exclude the compliance of the testing system. On the other hand, in terms of the fatigue behavior of unnotched specimens, the normalized S-N curves of unnotched [+45/90/$-$45/0]$_{s}$ and [0/90]$_{2s}$ specimens under tensile fatigue loading conditions are plotted in Figure \ref{fig:unnotched}a. As can be noted from the figure, the slope of S-N curve for the cross-ply specimen is slightly larger than the one of the quasi-isotropic specimen. The larger slope of S-N curve reflects the specimen with better resistance to fatigue loading. This is consistent with the results on the normalized stiffness degradation curves, as shown in Figure \ref{fig:unnotched}b, in which the structural stiffness of the cross-ply specimen deteriorates less significantly in comparison with the one of the quasi-isotropic specimen. The structural stiffness is defined as $K=(P_{peak}-P_{m})/(u_{peak}-u_{m})$ where $P_{peak}$ is the peak load, $P_{m}$ is the mean load, $u_{peak}$ is the displacement at the peak load for each cycle, and $u_{m}$ is the displacement at the mean load for each cycle. However, the endurance limit, generally considered as 2 million cycles, was achieved when the peak load in applied cyclic load reached roughly 30\%-35\% of the critical quasi-static load for both unnotched specimens.

\subsection{Multi-axial quasi-static tests on notched specimens}
\label{multiaxialquasistatictests}
\begin{figure} [H]
\center
\includegraphics[scale=0.47]{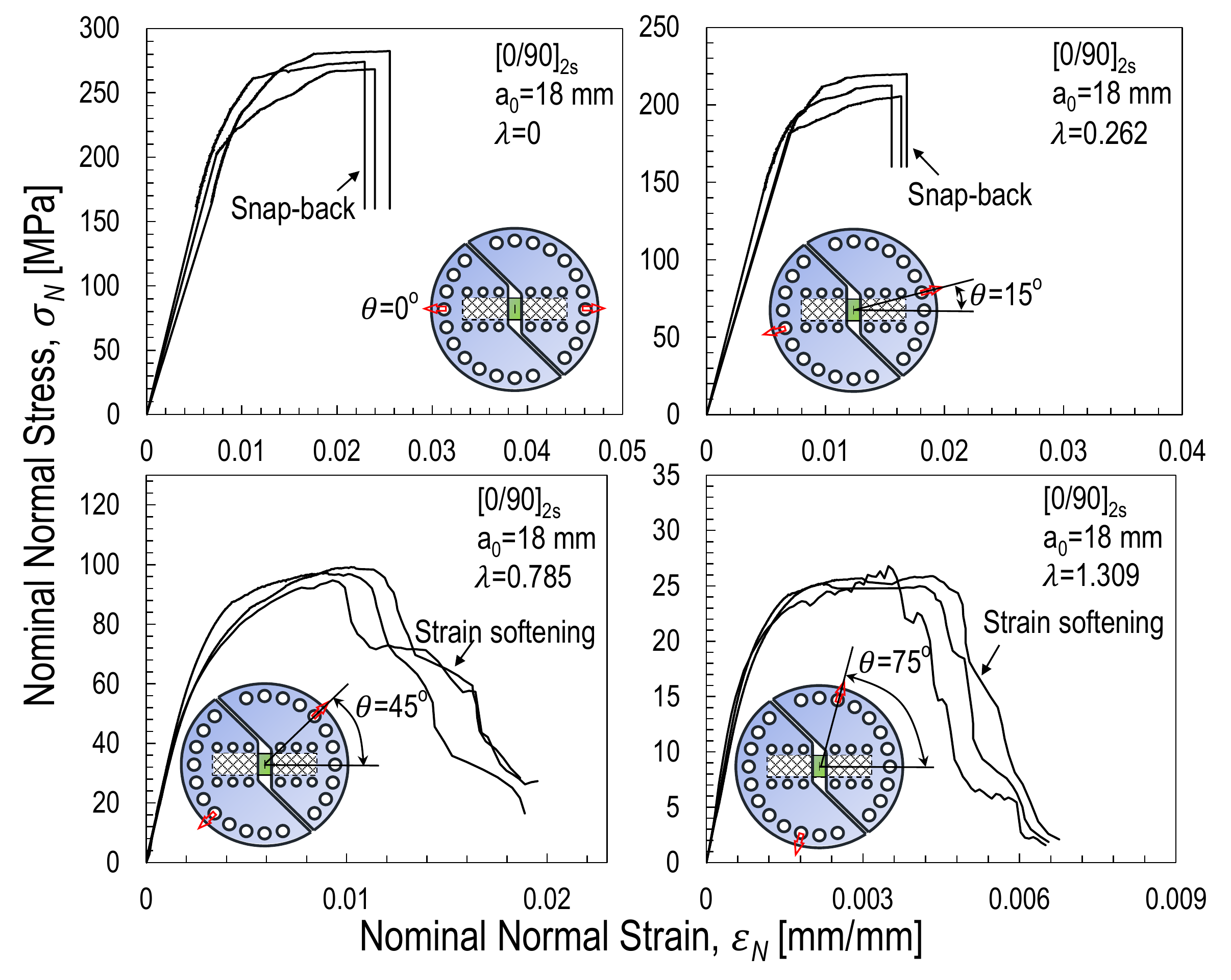}
\caption{Nominal normal stress vs. nominal normal strain obtained from the multi-axial quasi-static tests on the $[0/90]_{2s}$ specimen weakened by a 18 mm central crack.}
\label{fig:quasistaticcrossplynormal}
\end{figure}

\begin{figure} [H]
\center
\includegraphics[scale=0.47]{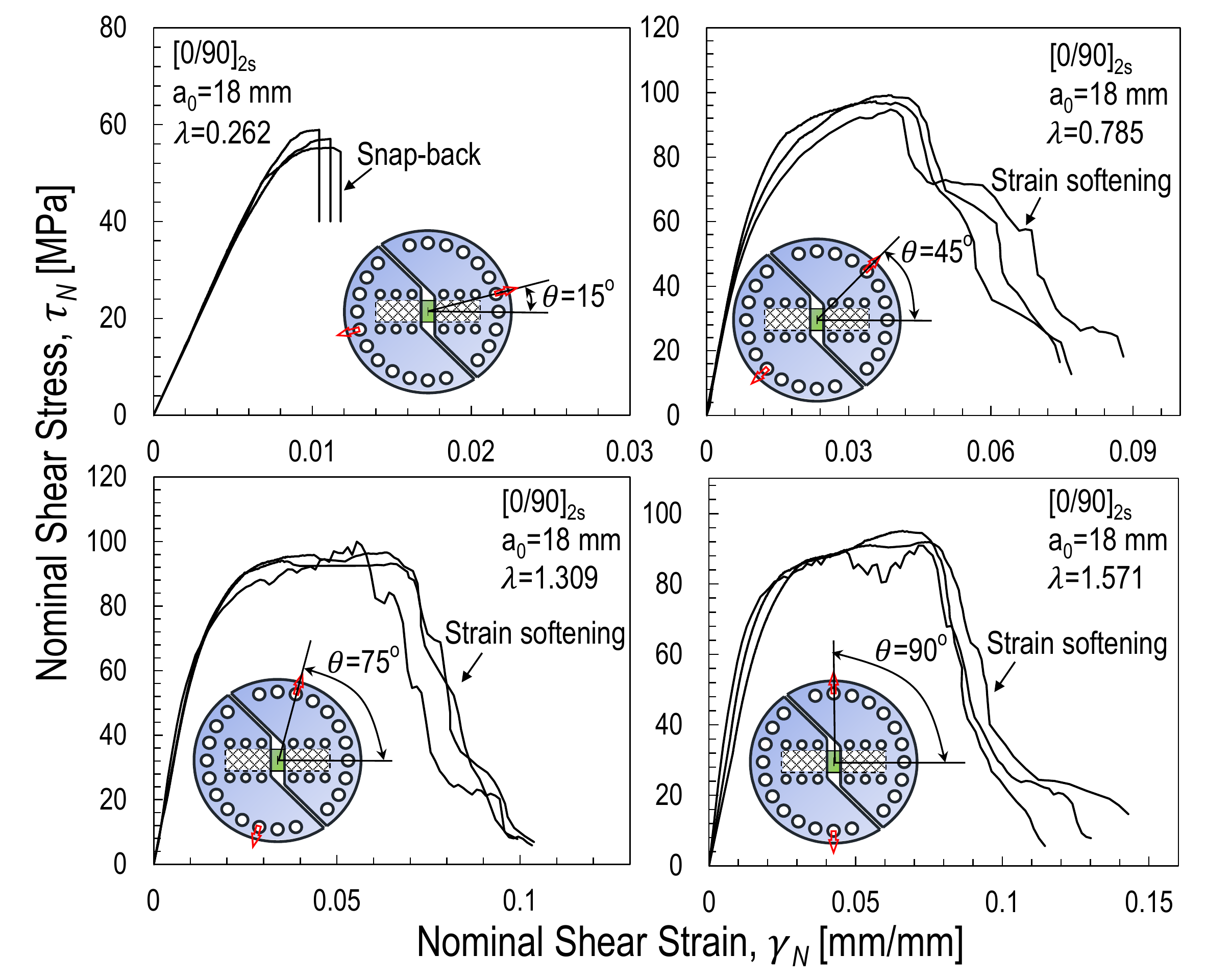}
\caption{Nominal shear stress vs. nominal shear strain obtained from the multi-axial quasi-static tests on the $[0/90]_{2s}$ specimen weakened by a 18 mm central crack.}
\label{fig:quasistaticcrossplyshear}
\end{figure}

\begin{figure} [H]
\center
\includegraphics[scale=0.46]{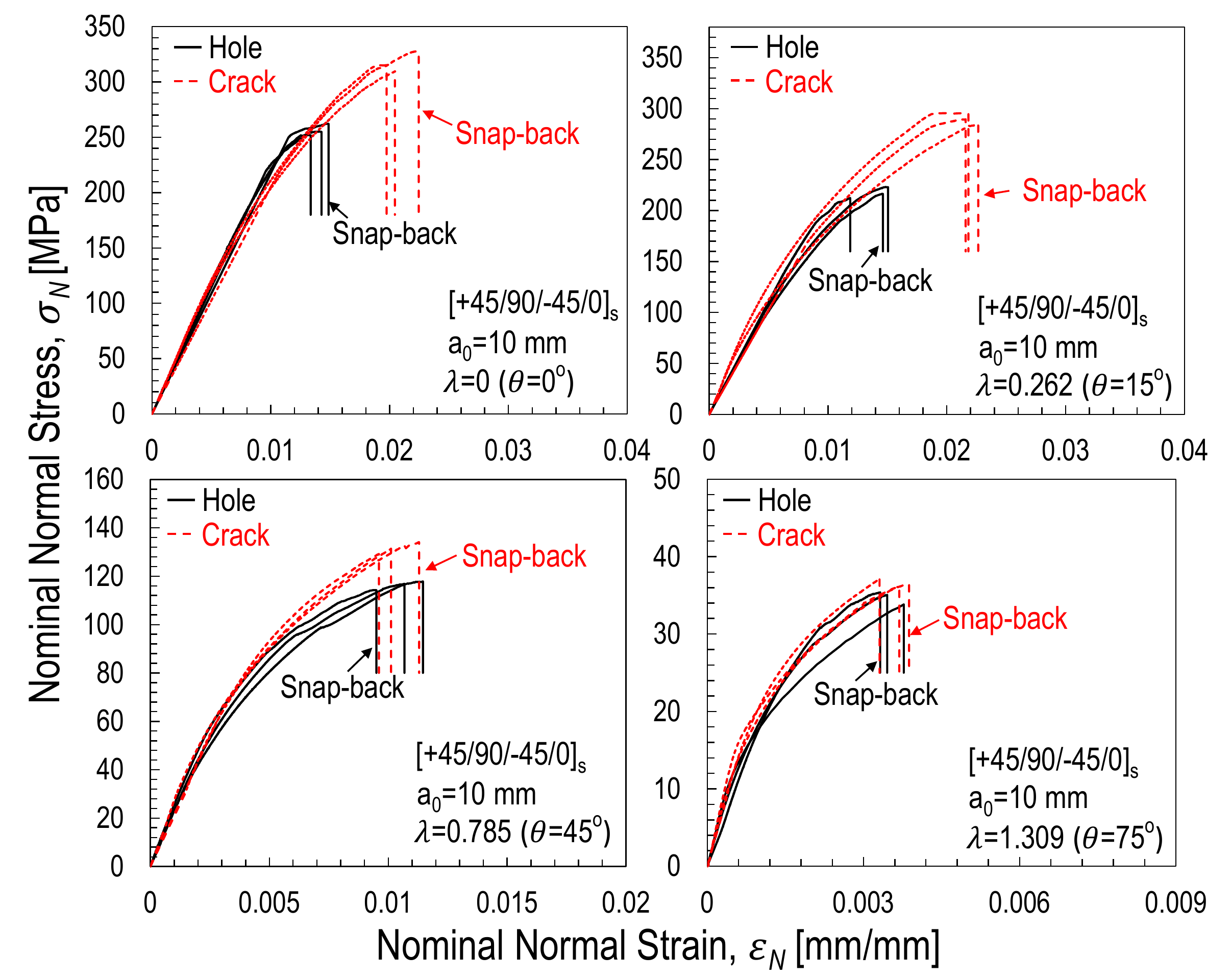}
\caption{Nominal normal stress vs. nominal normal strain obtained from the multi-axial quasi-static tests on the $[+45/90/-45/0]_{s}$ specimens weakened by a 10 mm central crack and a 10 mm hole respectively. All the tests summarized in this figure exhibited snap-back instability.}
\label{fig:quasistaticquasiisonormal}
\end{figure}

\begin{figure} [H]
\center
\includegraphics[scale=0.46]{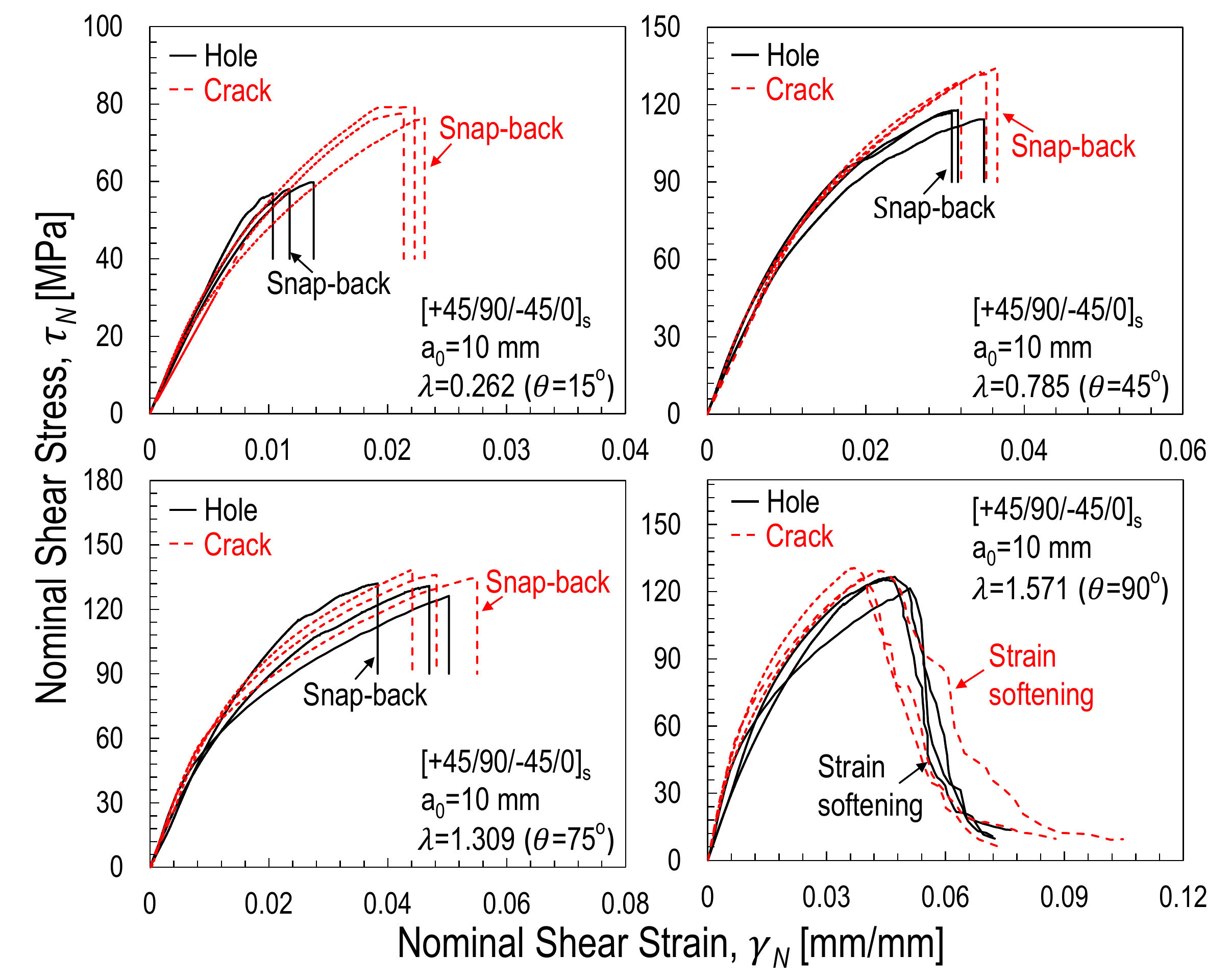}
\caption{Nominal shear stress vs. nominal shear strain obtained from the multi-axial quasi-static tests on the $[+45/90/-45/0]_{s}$ specimens weakened by a 10 mm central crack and a 10 mm hole respectively.}
\label{fig:quasistaticquasishear}
\end{figure}

The nominal stress and strain curves obtained from the multi-axial quasi-static tests are plotted in Figures \ref{fig:quasistaticcrossplynormal}-\ref{fig:quasistaticquasishear} for the following three specimen configurations: (1) [0/90]$_{2s}$ layup with a 18 mm central crack, (2) [+45/90/$-$45/0]$_{s}$ layup with a 10 mm hole, and (3) [+45/90/$-$45/0]$_{s}$ layup with a 10 mm central crack. In these figures, the normal and shear stresses are the nominal stresses in the net section, $\sigma_{N} = P$cos$\theta/[(w-a_0)t]$ and $\tau_{N} = P$sin$\theta/[(w-a_0)t]$, whereas the strain is calculated based on the difference in the displacements at two ends of the gauge area of the specimen obtained from Digital Imaging Correlation (DIC).

As can be noted from Figures \ref{fig:quasistaticcrossplynormal}-\ref{fig:quasistaticquasishear}, the stress-strain curves under multi-axial quasi-static loading are characterized by a significant non-linear behavior when the specimens are subjected to a combination of tension and shear, regardless of the layup. This phenomenon becomes more and more significant with increasing multiaxiality ratios due to the emergence of diffused, sub-critical matrix micro-cracking, splitting and delamination. These damage mechanisms contribute to the dissipation of the strain energy stored in the specimen inducing a significant non-linearity before reaching the failure load. A similar mechanism was reported in uniaxial tests on unidirectional laminates and two dimensional textile composites \cite{waas, sal1, kirane}. The mechanical behavior after peak stress is characterized by catastrophic failure due to snap-back instability \cite {Baz4,sal2} for all the investigated layups when the specimens are subjected to tension-dominated loading. On the other hand, the failure behavior becomes more and more stable with increasing the shear load component. Eventually, as shown in Figures \ref{fig:quasistaticcrossplynormal}-\ref{fig:quasistaticquasishear}, the post-peak behavior becomes stable and strain softening becomes evident. This is an indication of more pronounced quasi-brittleness with increasing the shear load component, a phenomenon even more pronounced for the $[+45/90/-45/0]_{s}$ specimens. It is worth mentioning here that the snap-back instability in tension-dominated loads is a structural, not a material phenomenon. To explore the post-peak behavior, one could leverage the new device proposed by the authors \cite{patent}. However, this is beyond the scope of the present work.

\begin{figure} [H]
\center
\includegraphics[scale=0.46]{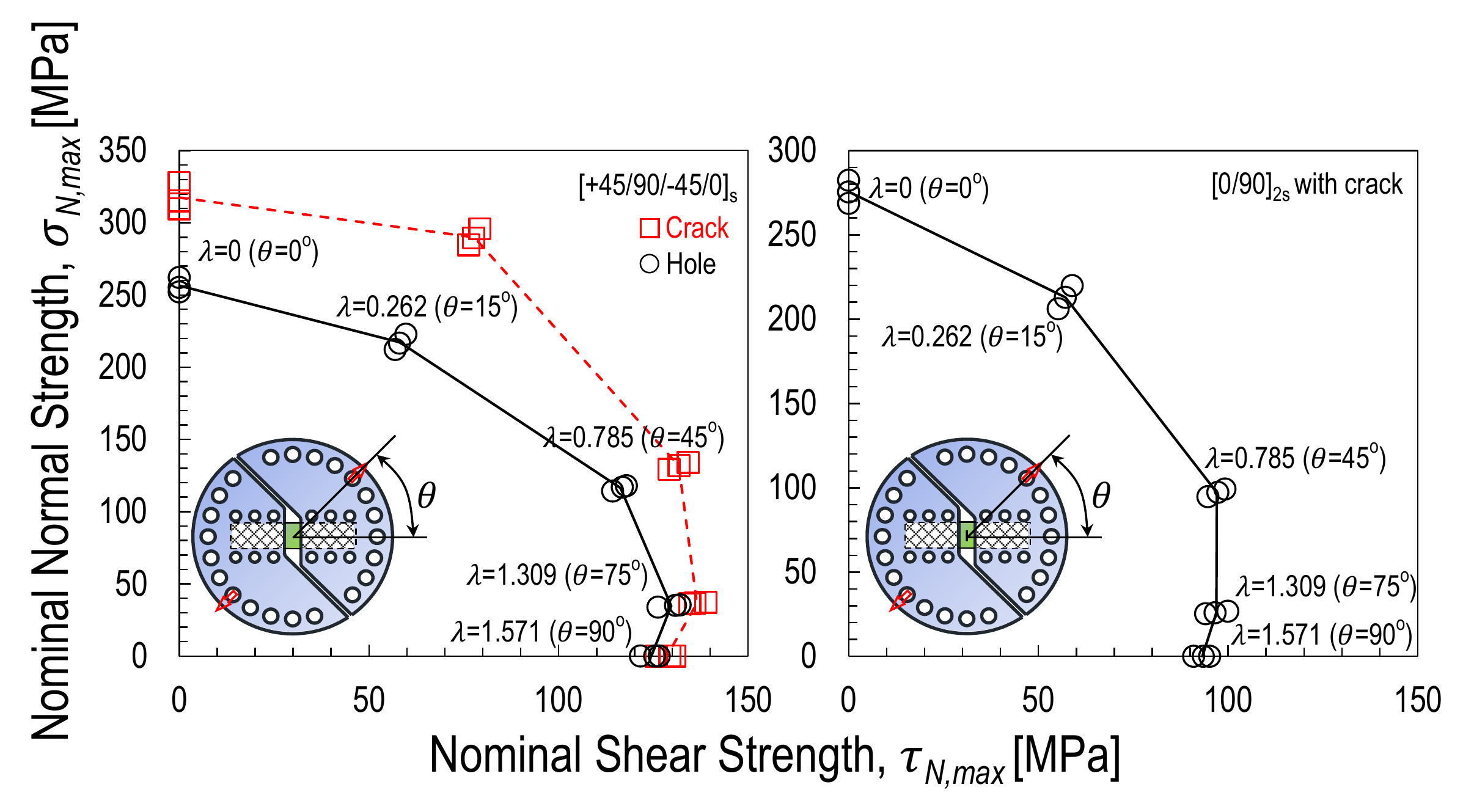}
\caption{Failure envelopes of notched $[0/90]_{2s}$ and $[+45/90/-45/0]_{s}$ specimens under multi-axial quasi-static loading conditions. Note that the nominal normal and shear strength are defined as $P_{max}cos\theta/[(w-a)t]$ and $P_{max}sin\theta/[(w-a)t]$ and the multiaxiality ratio is defined as $\lambda= arctan(\tau_{N}/\sigma_{N})$.}
\label{fig:failureenvelop}
\end{figure}

It is interesting to investigate the relationship between the normal and shear strength (critical net cross-section stress) as a function of the multiaxiality ratio. The failure envelopes for the foregoing notched specimens under multi-axial quasi-static tests are plotted in Figure \ref{fig:failureenvelop} and the details can be found in Table \ref{tab:notchedproperties}. In this plot, the normal and shear strengths are defined as $\sigma_{N,max}=P_{max}$cos$\theta/[(w-a_0)t]$ and $\tau_{N,max}=P_{max}$sin$\theta/[(w-a_0)t]$ respectively where $P_{max}$ is the critical load. As can be noted from Figure \ref{fig:failureenvelop}, the failure envelope of quasi-isotropic [+45/90/$-$45/0]$_{s}$ layup with a 10 mm central crack encompasses the one with a 10 mm open hole hinting the fact that the hole affects the structural strength more severely than a crack. A similar phenomenon, associated to the greater splitting development in load-bearing plies for the specimens with an intra-laminar central crack compared to the ones with an open hole, has been reported by several authors on notched Carbon Fiber Reinforced Polymer (CFRP) laminates \cite{Tan,Xu}. However, since the failure behavior of notched composites depends on the size of the non-linear Fracture Process Zone (FPZ) occurring in the presence of a large stress-free crack compared to the specimen width \cite{Baz1,Sal,Baz2,Baz3,Yao1,Yao2,Yao4,Seung1,Seung2,Deleo}, geometrically scaled specimens with an open hole or a central crack need to be investigated to better understand this phenomenon. The size effect tests are undergoing and will be presented in future publications. Another subject of future investigations is the evolution of the fracture toughness with the multiaxiality ratio which has been studied in nanocomposites \cite{Zap,Samit}.

\subsection{Multi-axial fatigue tests on notched specimens}
\subsubsection{Evolution of structural stiffness vs. multiaxiality ratio}
\label{sec:evolutionofstiffness}
\begin{figure} [H]
\center
\includegraphics[scale=0.4]{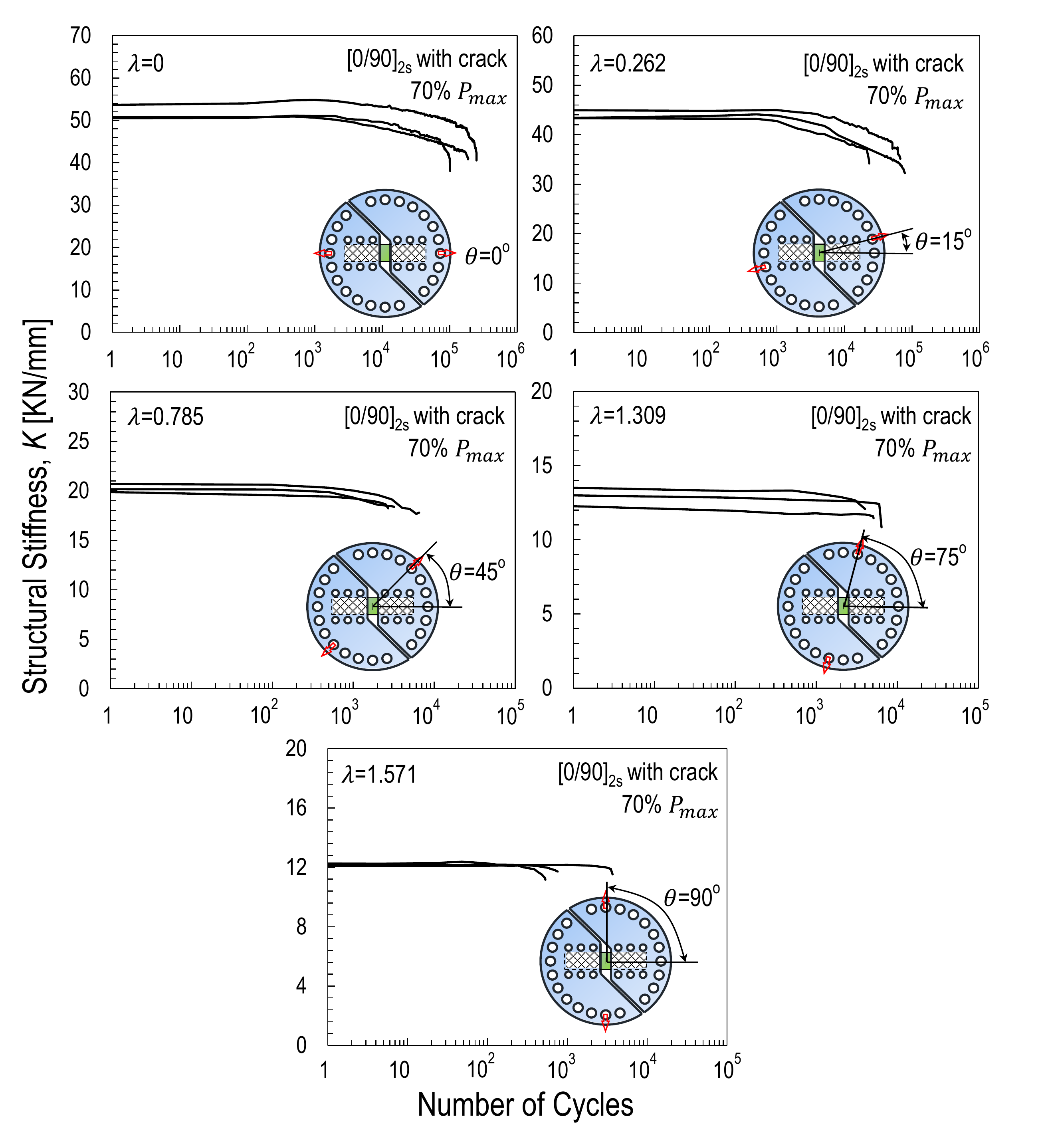}
\caption{Evolution of structural stiffness vs. number of cycles measured during the multi-axial fatigue tests (70\% $P_{max}$) on the $[0/90]_{2s}$ specimen with a 10 mm central crack.}
\label{fig:stiffnessdegradationcrossply70}
\end{figure}

\begin{figure} [H]
\center
\includegraphics[scale=0.55]{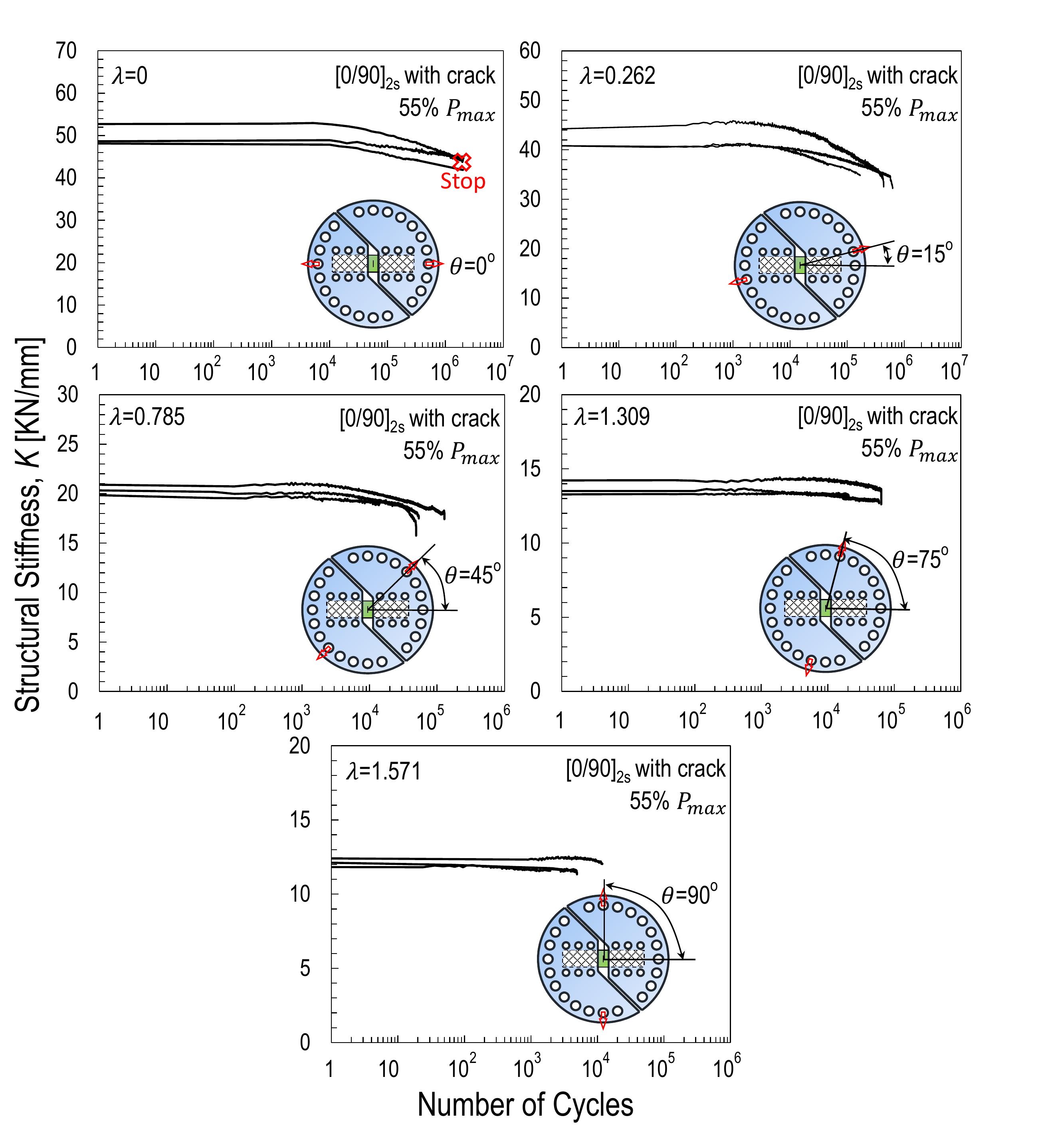}
\caption{Evolution of structural stiffness vs. number of cycles measured during the multi-axial fatigue tests (55\% $P_{max}$) on the  $[0/90]_{2s}$ specimen with a 10 mm central crack.}
\label{fig:stiffnessdegradationcrossply55}
\end{figure}

\begin{figure} [H]
\center
\includegraphics[scale=0.55]{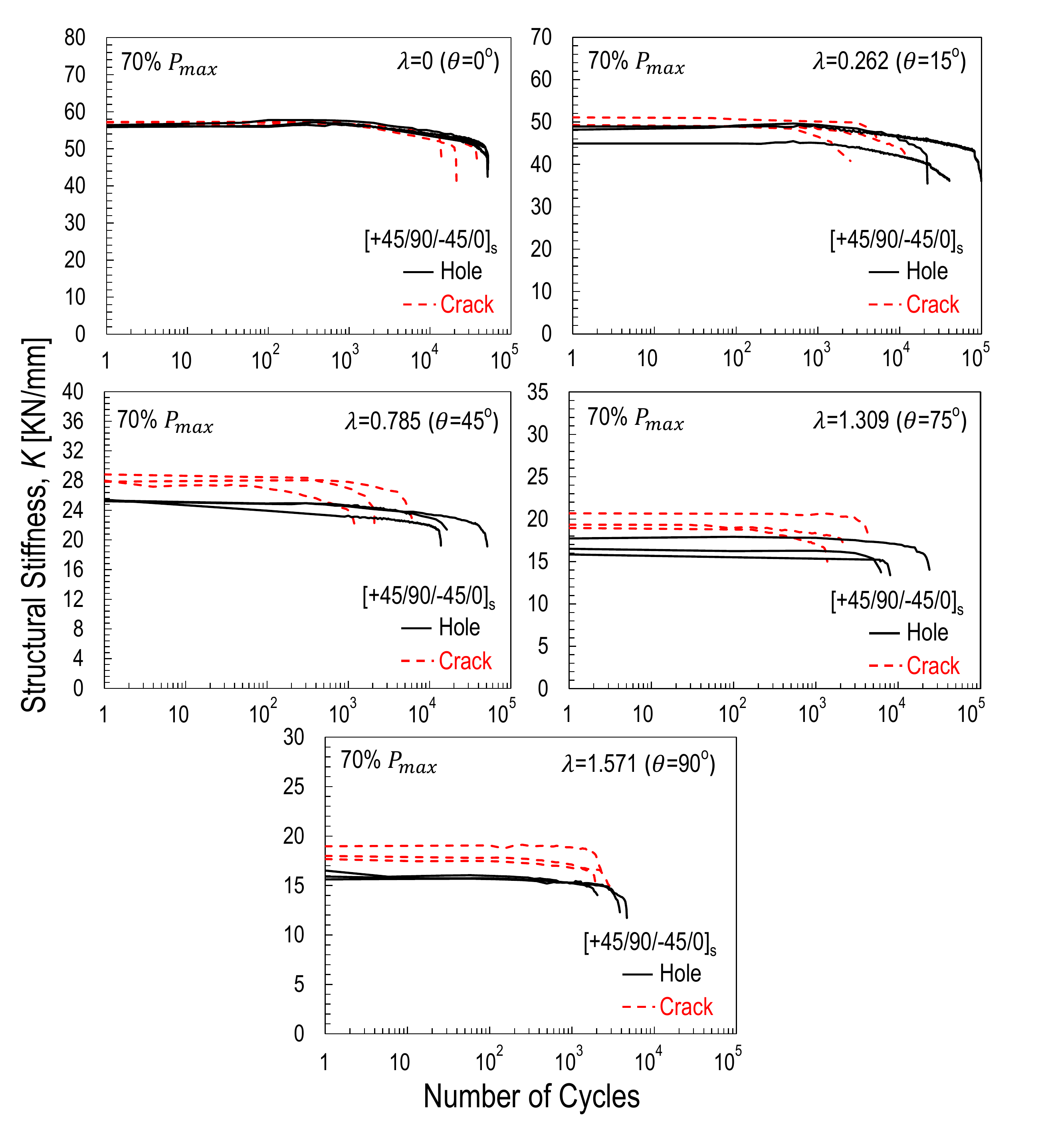}
\caption{Evolution of structural stiffness vs. number of cycles measured during the multi-axial fatigue tests (70\% $P_{max}$) on the $[+45/90/-45/0]_{s}$ specimens with a 10 mm central crack and a 10 mm hole.}
\label{fig:stiffnessdegradationisotropic70}
\end{figure}

\begin{figure} [H]
\center
\includegraphics[scale=0.37]{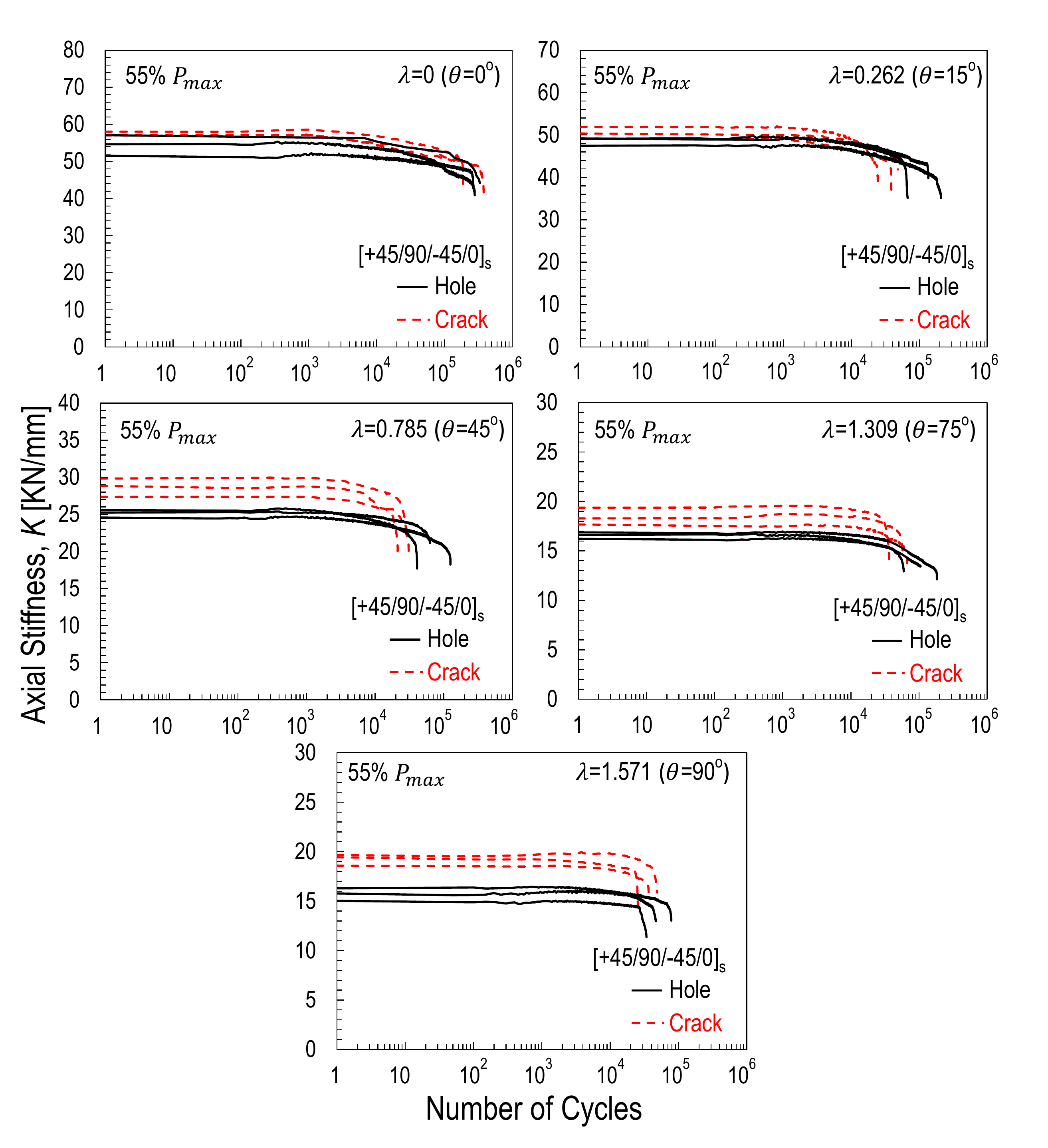}
\caption{Evolution of structural stiffness vs. number of cycles measured during the multi-axial fatigue tests (55\% $P_{max}$) on the $[+45/90/-45/0]_{s}$ specimens with a 10 mm central crack and a 10 mm hole.}
\label{fig:stiffnessdegradationisotropic55}
\end{figure}

Having discussed the failure behavior of notched laminates under multi-axial quasi-static loading, the fatigue failure behavior needs to be investigated. The evolution of the structural stiffness obtained from the multi-axial fatigue tests is represented in Figures \ref{fig:stiffnessdegradationcrossply70}-\ref{fig:stiffnessdegradationisotropic55} for all the studied notched specimens and two loading  cases (70\% and 55\% of $P_{max}$). In these plots, the stiffness has the same meaning defined for the analysis of unnotched specimens under tensile fatigue loading condition.

\begin{figure} [H]
\center
\includegraphics[scale=0.4]{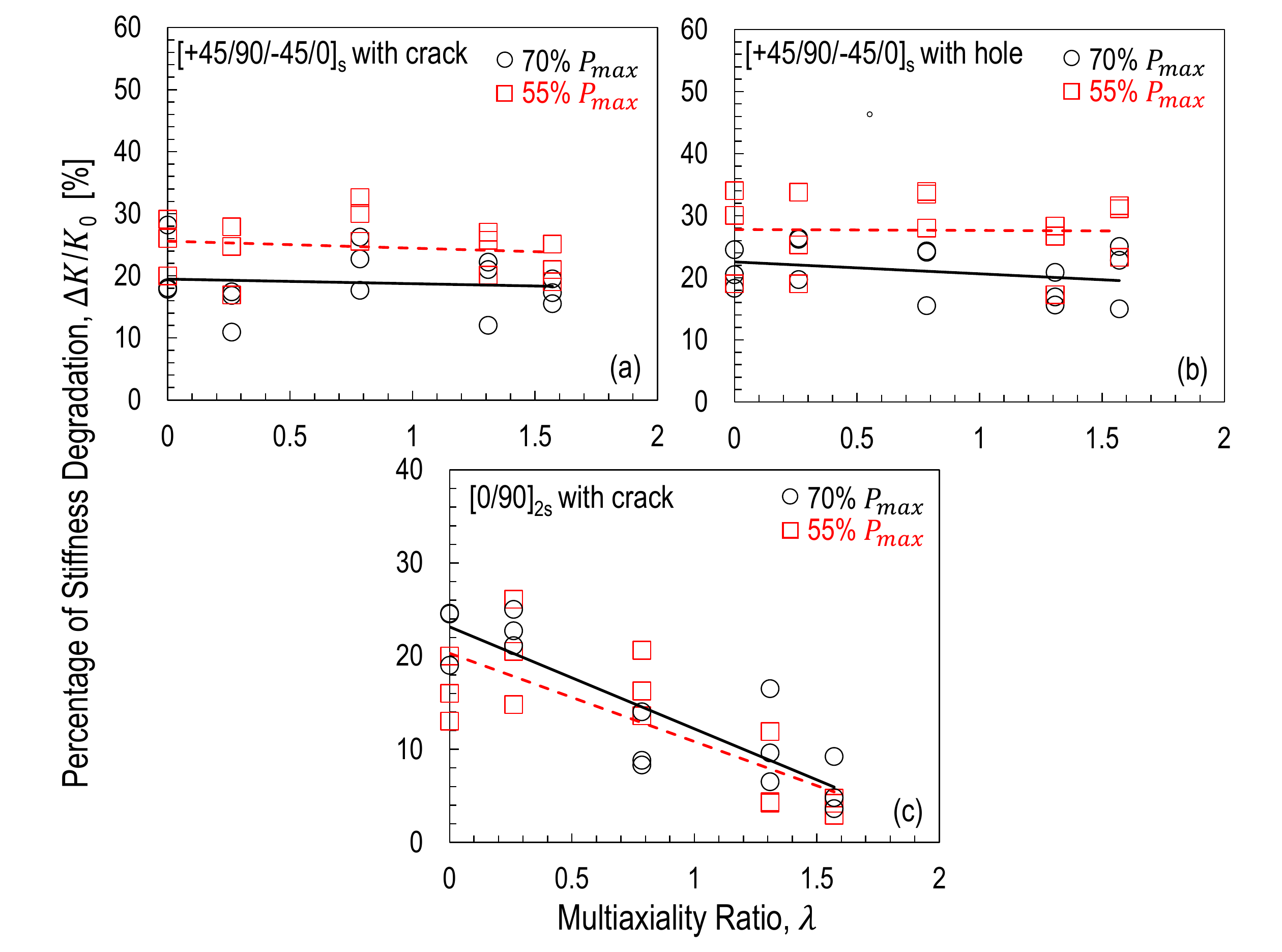}
\caption{The percentage of stiffness degradation right before catastrophic failure vs. multiaxiality ratio for notched quasi-isotropic and cross-ply specimens. Note that $K_{0}$ is the initial stiffness.}
\label{fig:degradationmultiaxialratio}
\end{figure}

\begin{figure} [H]
\center
\includegraphics[scale=0.75]{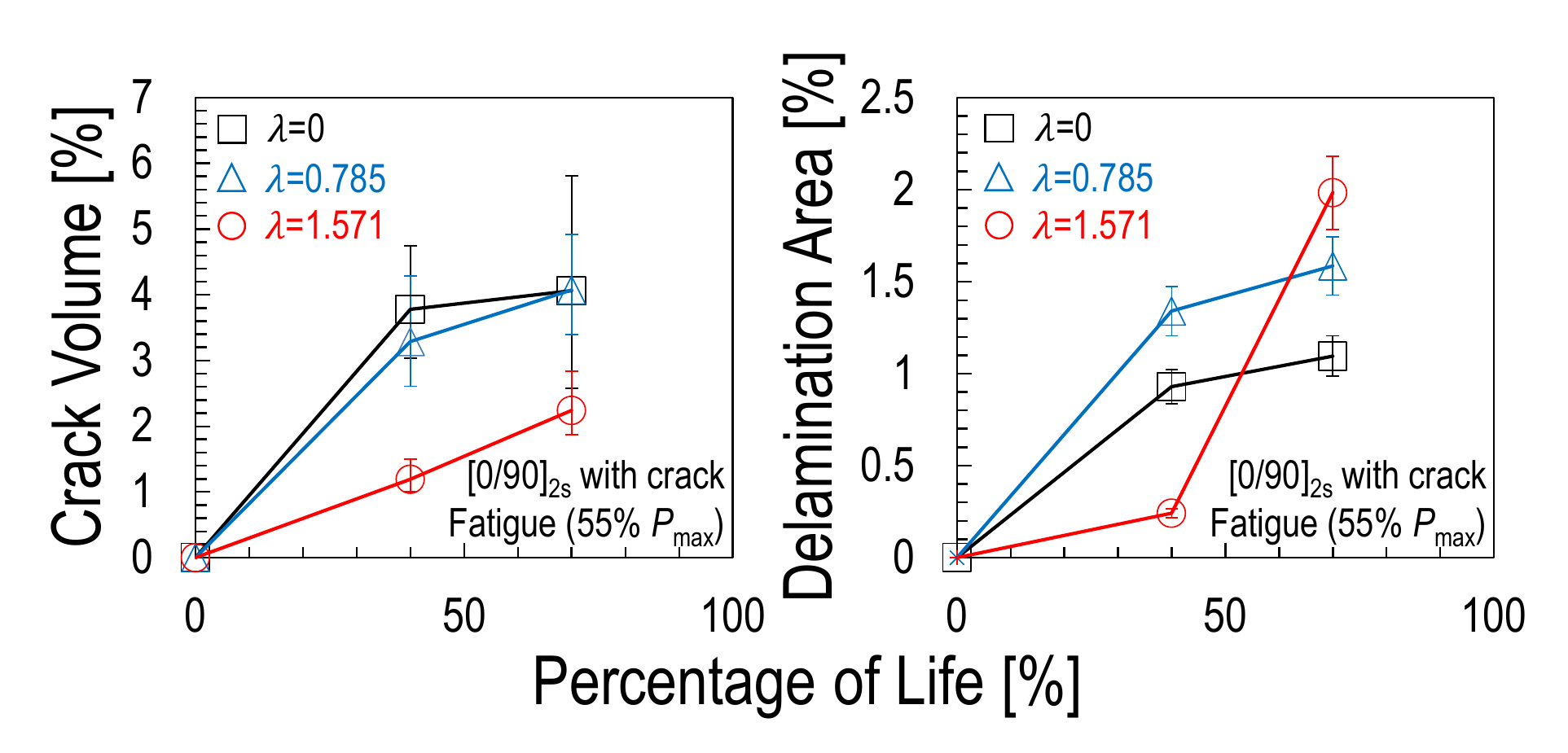}
\caption{The evolution of the total crack volume and delamination area for the $[0/90]_{2s}$ specimens weakened by a 18 mm central crack as a function of the percentage of fatigue life for three multiaxiality ratios. Results are based on micro-computed tomography.}
\label{fig:crossdamageanalysis}
\end{figure}

As can be noted from Figures \ref{fig:stiffnessdegradationcrossply70}-\ref{fig:stiffnessdegradationcrossply55}, for cross-ply [0/90]$_{2s}$ specimen with a central crack, the stiffness decreases of roughly 10\% in the early stage of tension-dominated fatigue tests ($\lambda=0, 0.262$) while no significant degradation can be observed throughout fatigue life of specimens subjected to shear-dominated fatigue loading. This is mainly due to the growth of transverse matrix cracking during tension-dominated fatigue loading \cite{talreja1}. Eventually, as illustrated in Figure \ref{fig:degradationmultiaxialratio}c, the stiffness decreases of about 20\% before catastrophic failure for multiaxiality ratio $\lambda=0$ whereas only 5\% degradation in stiffness was observed for multiaxiality ratio $\lambda=1.571$. This can be attributed to a significant reduction of the splitting in $90^{\circ}$ plies under shear-dominated loading compared to tension-dominated loading before catastrophic failure as shown in Figure \ref{fig:crossplydamage} through the quantitative damage analysis by using X-ray micro-computed tomography ($\mu$-CT). Despite this reduction leading to a lower total crack volume for shear-dominated loading, the final failure is eventually triggered by the rapid growth of delamination in the late stage of fatigue tests as shown in Figure \ref{fig:crossdamageanalysis}. 

\begin{figure} [H]
\center
\includegraphics[scale=0.65]{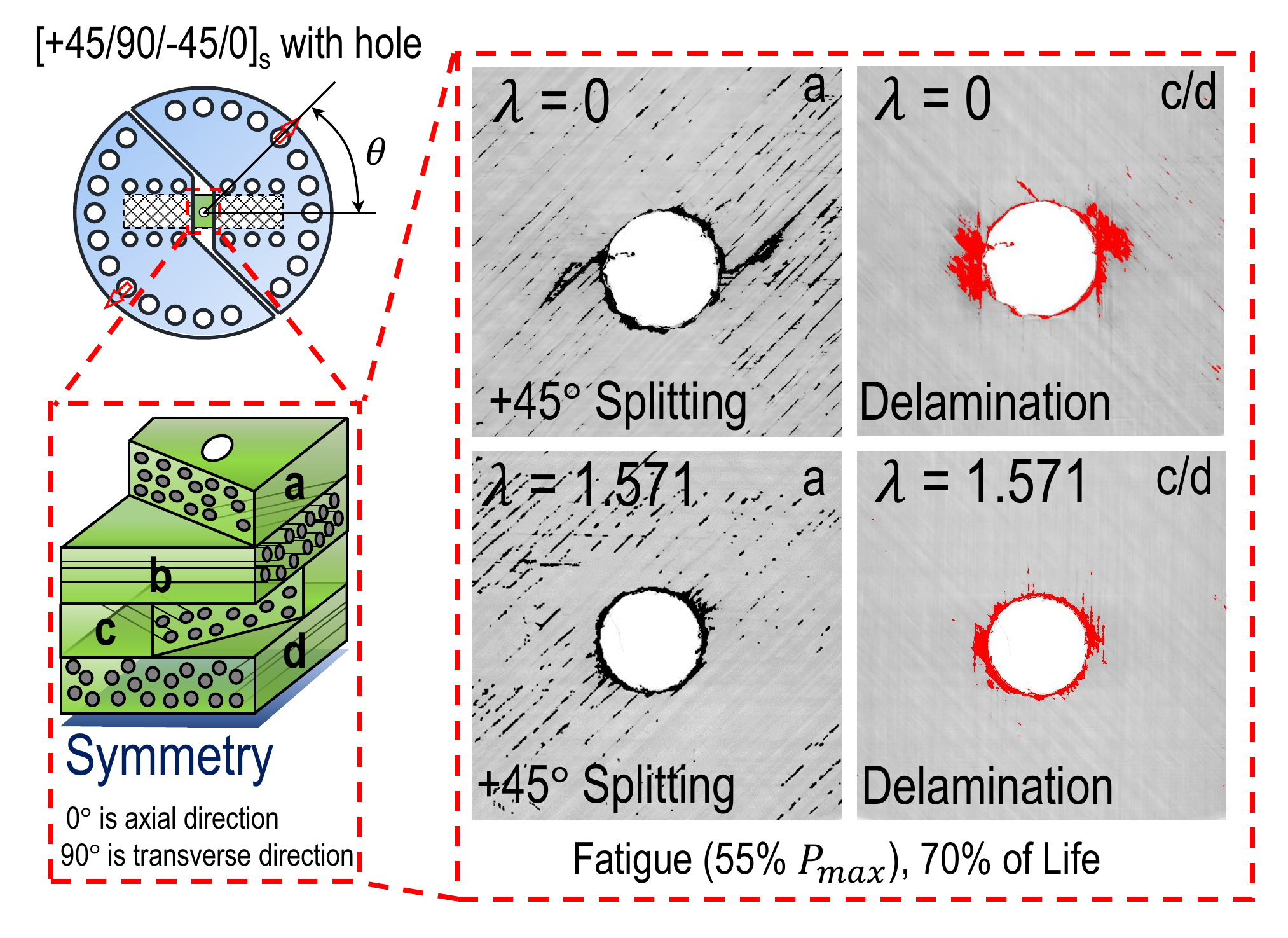}
\caption{Analysis of main fatigue damage mechanisms before final failure by micro-computed tomography. Comparison on fatigue damage in [+45/90/$-$45/0]$_{s}$ specimens weakened by a 10 mm open hole for multiaxiality ratio $\lambda=0$ and $\lambda=1.571$. Note that the original colors of the images were inverted for a better visualization on the damage.}
\label{fig:quasidamage}
\end{figure}

 However, this is not the case for quasi-isotropic [+45/90/$-$45/0]$_{s}$ specimen with an open hole or a central crack since in both cases a gradual stiffness degradation throughout the fatigue life can be noted for various multiaxiality ratios. In contrast to cross-ply [0/90]$_{2s}$ specimens, the stiffness of quasi-isotropic specimens with an open hole or a central crack deteriorates roughly 19\% to 25\% for all the investigated multiaxiality ratios as shown in Figures \ref{fig:degradationmultiaxialratio}a-b. This consistent deterioration is mainly due to the significant development of splitting and delamination before catastrophic failure for all the multi-axial tests as shown in Figure \ref{fig:quasidamage} exemplifying the typical damage in specimens featuring an open hole. Similar damage was observed in specimens featuring a central crack.

\begin{figure} [H]
\center
\includegraphics[scale=0.75]{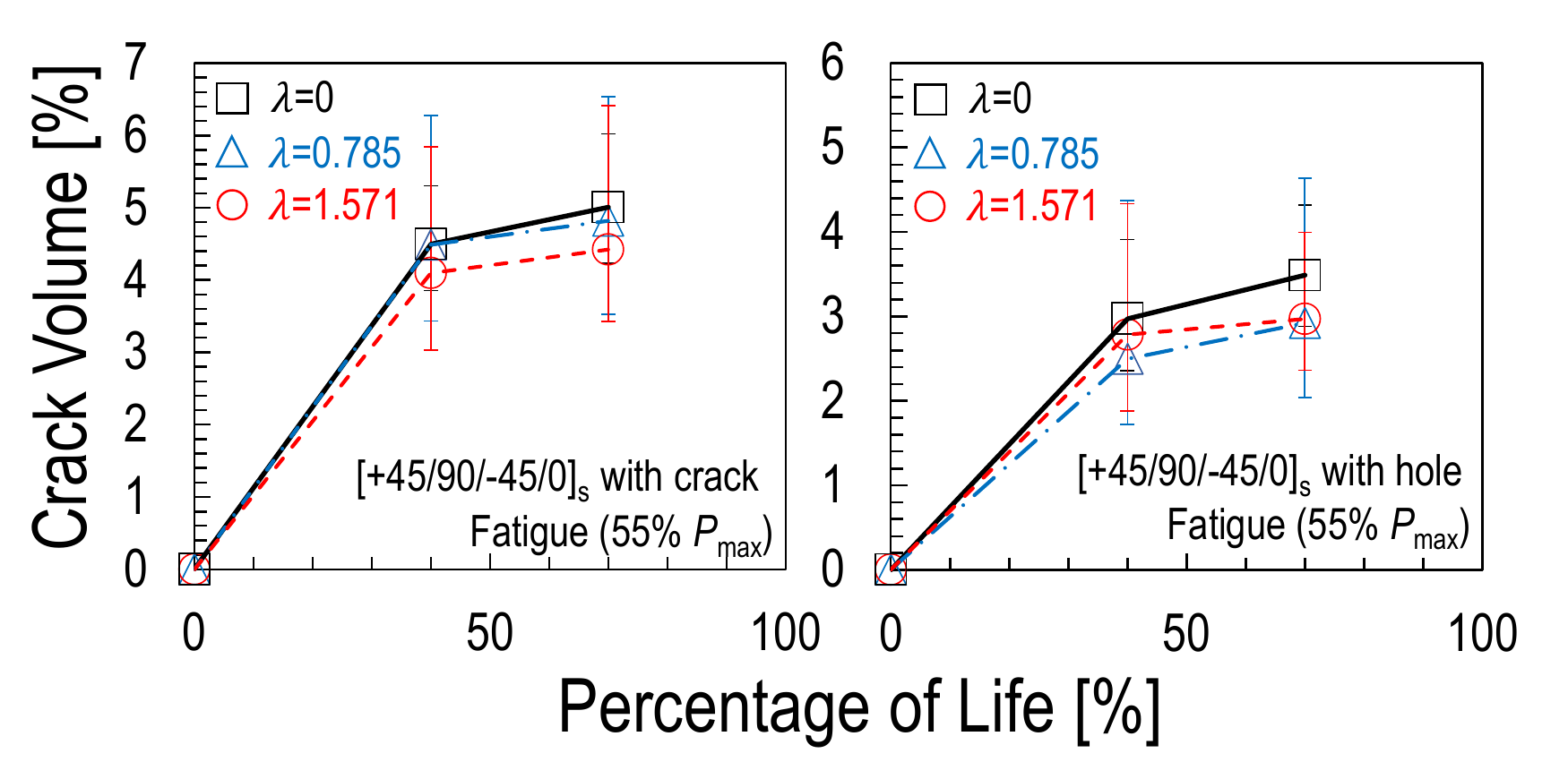}
\caption{The evolution of the total crack volume for notched [+45/90/$-$45/0]$_{s}$ specimens as a function of the percentage of fatigue life for three multiaxiality ratios. The graphs compare hole and crack results obtained from micro-computed tomography.}
\label{fig:quasidamageanalysis}
\end{figure}

In addition to the foregoing observations, it can be noted from Figures \ref{fig:stiffnessdegradationisotropic70}-\ref{fig:stiffnessdegradationisotropic55} that the fatigue performance of quasi-isotropic [+45/90/$-$45/0]$_{s}$ specimens with a central crack is generally worse than the one of the same layup featuring an open hole. As shown in Figure \ref{fig:quasidamageanalysis}, the quantitative damage analysis by using $\mu$-CT technique confirms that the worse performance of the quasi-isotropic specimens weakened by a central crack compared to the ones weakened by a central hole is associated to a significant larger amount of crack volume. It is interesting to note that this is in contrast to the quasi-static results which pointed to the central hole case as being the most critical. This structural phenomenon is due to a different evolution of the Fracture Process Zone (FPZ) in quasi-static regime compared to fatigue. There is no doubt that the process of FPZ development is strongly affected by the size of the specimens (width, notch size, etc.). This size effect deserves further studies and it will be the subject of future publications by the authors.

\subsubsection{S-N curve vs. multiaxiality ratio}
\begin{figure} [H]
\center
\includegraphics[scale=0.45]{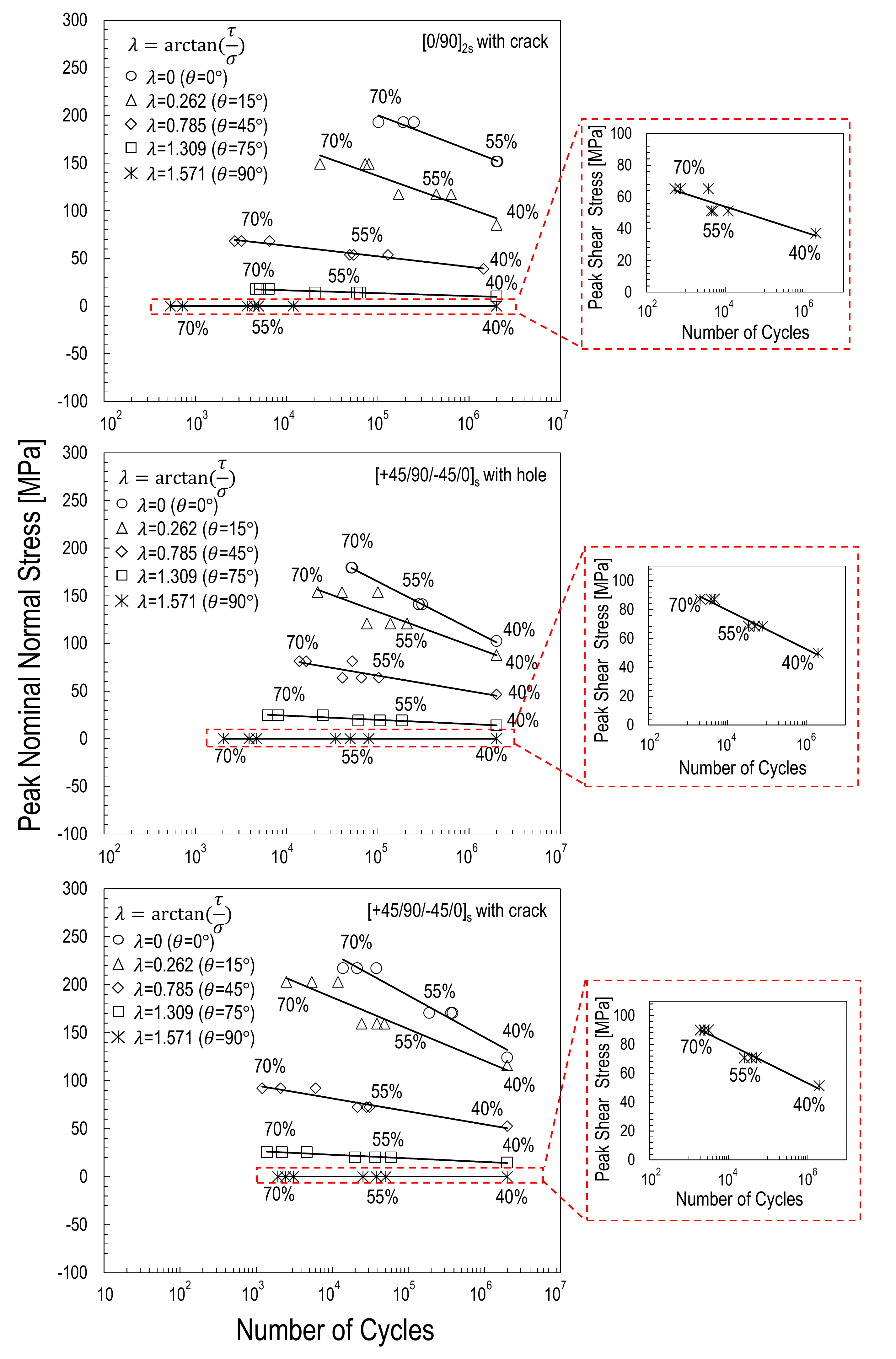}
\caption{S-N curves vs. multiaxiality ratio measured from the multi-axial fatigue tests on notched quasi-isotropic and cross-ply specimens.}
\label{fig:SNcurves}
\end{figure}

To gain a better understanding of the fatigue behavior of notched laminates under various multiaxiality ratios, S-N curves were constructed by leveraging the fatigue tests with three different loading cases (70\%, 55\% and 40\% of $P_{max}$). As illustrated in Figure \ref{fig:SNcurves}, the peak nominal normal stress is plotted as a function of number of cycles in the semi-logarithmic coordinate for notched specimens. In these plots, the slopes of S-N curves decrease as multiaxiality ratio increases which indicates that the deterioration of the fatigue behavior happens severely when the loading condition transits from tension to shear. This phenomenon is consistent with the previous experimental investigations on both tabular and cruciform specimens featuring a circular hole \cite{quaresimin8,francis,jones,Fuji5,Fuji6,Fuji7}. It is worth mentioning here that, for specimens under the fatigue loading of pure shear ($\lambda=1.571$), the slope of S-N curve is zero due to the definition of the multiaxiality ratio. Thus, additional S-N curves are plotted in Figure \ref{fig:SNcurves} for these specific cases. 

\begin{figure} [H]
\center
\includegraphics[scale=0.65]{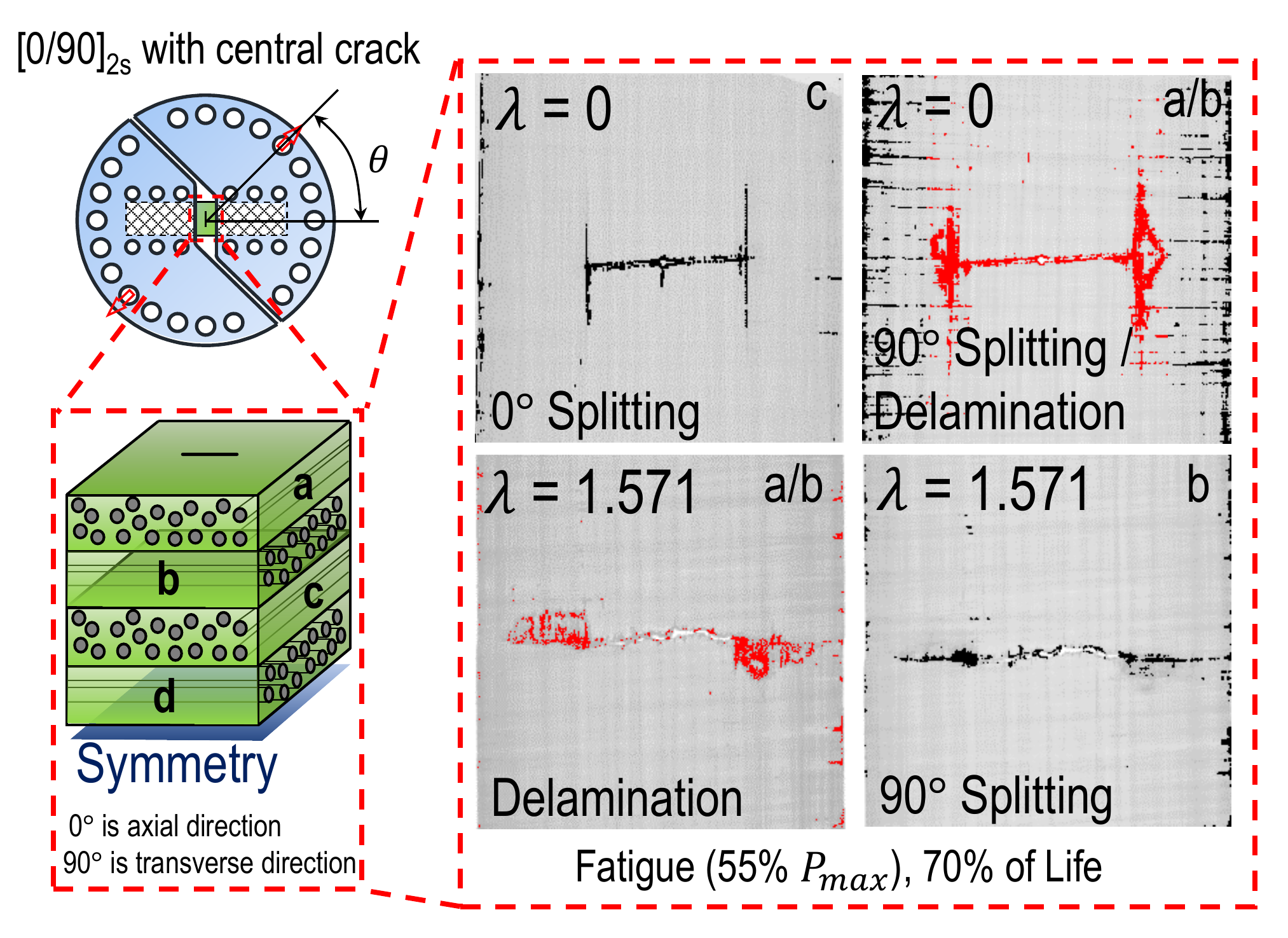}
\caption{Analysis of main fatigue damage mechanisms before final failure by micro-computed tomography. Comparison on fatigue damage in $[0/90]_{2s}$ specimens weakened by a 18 mm central crack for multiaxiality ratio $\lambda=0$ and $\lambda=1.571$. Note that the original colors of the images were inverted for a better visualization on the damage.}
\label{fig:crossplydamage}
\end{figure}

\begin{figure} [H]
\center
\includegraphics[scale=0.55]{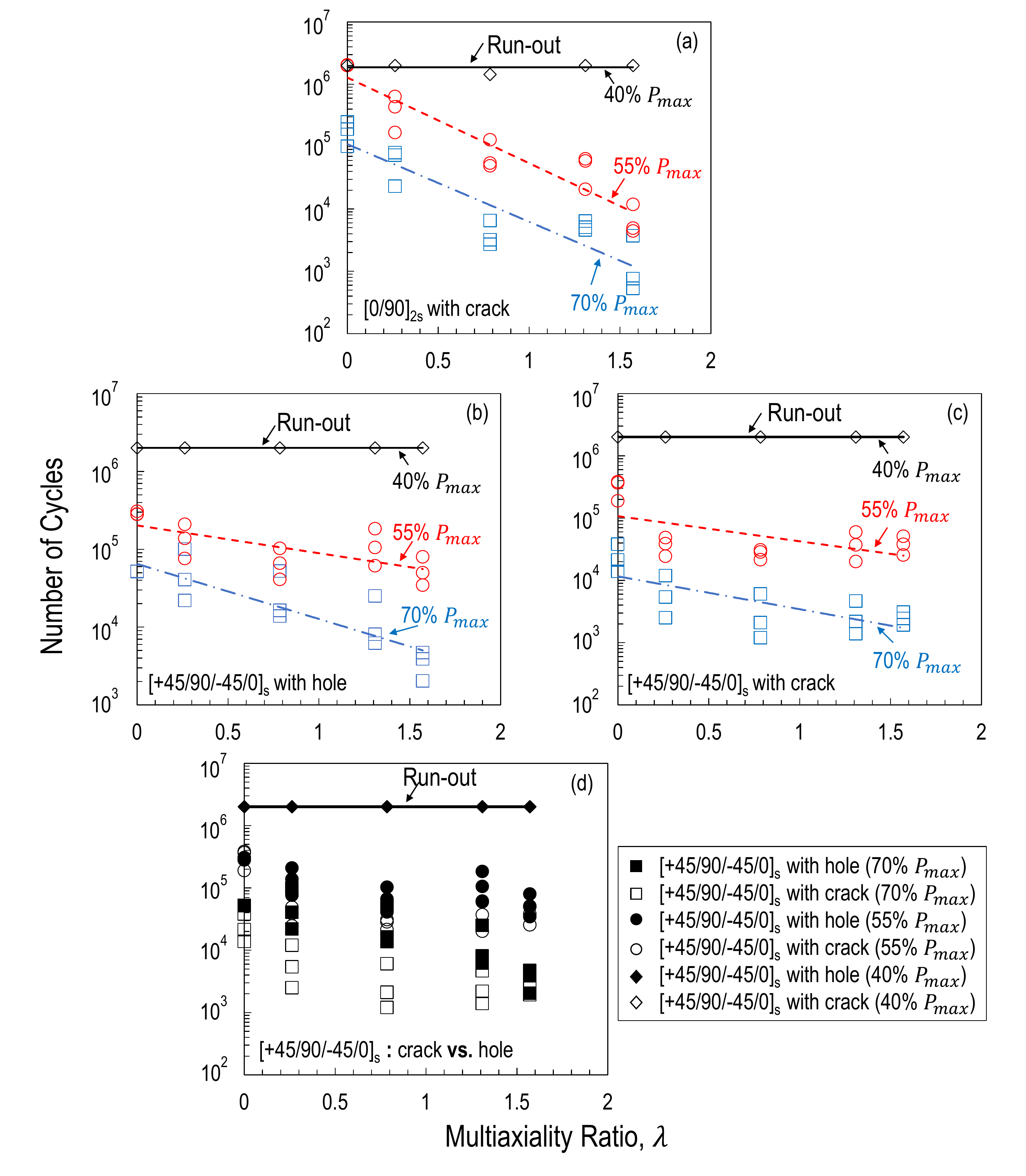}
\caption{Number of cycles to failure vs. multiaxiality ratio for notched quasi-isotropic and cross-ply specimens.}
\label{fig:fatiguelifevsmultiaxialratio}
\end{figure}

As far as the fatigue life is concerned, as illustrated in Figure \ref{fig:fatiguelifevsmultiaxialratio} and tabulated in Table \ref{tab:fatiguetimelifes}, the number of cycles to failure decreases with increasing the multiaxiality ratio for both two fatigue loading conditions (70\% and 55\% of $P_{max}$). This cause the decrease of the slopes of S-N curves as mentioned in the forgoing discussion. Furthermore, it was shown in Figure \ref{fig:fatiguelifevsmultiaxialratio}a that the endurance limit (2 million cycles) of notched cross-ply [0/90]$_{2s}$ specimens under the fatigue loading of pure tension can be achieved approximately when 55\% of $P_{max}$ is applied as a peak load in the cyclic load. This is much higher than the condition for the forgoing unnotched specimens featuring the same layup under tensile fatigue loading condition. This is probably due to the reduced stress concentration of a central crack caused by splitting in $0^{\circ}$ plies occurring at the notch tip during the fatigue \cite{kortschot,spearing,morais}. However, the fatigue loading condition for the endurance limit of notched cross-ply [0/90]$_{2s}$ specimens at multiaxiality ratios except for pure tension is about 40\%  of $P_{max}$, the similar condition for notched quasi-isotropic [+45/90/$-$45/0]$_{s}$ specimens at all the investigated multiaxiality ratios in order to reach the endurance limit during the fatigue. This lower limit is close to the cases of investigated unnotched quasi-isotropic and cross-ply specimens under tensile fatigue loading condition which indicates that the effect of the splitting in those cases is mitigated and does not help to significantly reduce the stress concentration at the notch tip thus not improving the fatigue resistance. In fact, the splitting in $0^{\circ}$ plies occurring at the notch tip for those cases is not a dominant mechanism as shown in Figure \ref{fig:crossplydamage}. The main damage mechanism transits from the significant splitting in $0^{\circ}$ plies to the delamination between $0^{\circ}$ and $90^{\circ}$ plies when the multiaxiality ratio increases. More detailed information on the fracturing morphology of notched quasi-isotropic and cross-ply laminates under multi-axial quasi-static and fatigue loading will be discussed in a future publication focusing on the damage mechanisms and the quantitative damage analysis by using X-ray micro-computed tomography ($\mu$-CT). 

\section{Conclusions}
This work investigated the failure behavior of notched quasi-isotropic and cross-ply laminates under both multi-axial quasi-static and fatigue loading. Based on the results obtained in this study, the following conclusions can be elaborated:

1. for notched laminates under multi-axial quasi-static loading, the stress-strain behavior is characterized by a significant non-linearity. This phenomenon becomes more and more significant with increasing shear load components due to the emergence of diffused, sub-critical matrix micro-cracking and delamination. This conclusion is supported by micro-computed tomography analysis.

2. The multiaxiality ratio influences significantly also the post-peak behavior. Catastrophic failure due to snap-back instability occurs when the specimens are subjected to tension-dominated loading whereas the post-peak behavior becomes stable and strain softening becomes evident in the case of shear-dominated loading;

3. Leveraging the multi-axial data, failure envelops in terms of peak nominal normal and shear stresses were constructed. Interestingly, the quasi-isotropic laminates featuring an open hole of diameter $a_0$ exhibited a lower strength compared to specimens weakened by a central crack with length equal to $a_0$. This phenomenon depends on the evolution of the Fracture Process Zone (FPZ) as a function of the multiaxiality ratio and specimen size. Further size effect tests are undergoing to better clarify this phenomenon.

4. the forgoing quasi-static data are in contrast to the results of the fatigue tests showing the central crack case being the most critical. This structural phenomenon may be due to a different evolution of the Fracture Process Zone (FPZ) in quasi-static regime compared to fatigue. This aspect is particularly important for structural design since it shows that failure under quasi-static loading and fatigue can occur by a completely different damage progression. The efficient fatigue design of composite structures demands the formulation of damage models that can capture such evolution in the context of the loading configuration, stacking sequence, and structure size and geometry.

5. the evolution of the structural stiffness degradation in notched quasi-isotropic and cross-ply laminates is substantially different. While the structural stiffness exhibits a similar amount of degradation before catastrophic failure for quasi-isotropic laminates at different multiaxiality ratios, the stiffness of cross-ply laminates deteriorates about 20\% before catastrophic failure for a multiaxiality ratio $\lambda=0$ but only 5\% degradation for a multiaxiality ratio $\lambda=1.571$. This is mainly due to a significant reduction of transverse matrix cracking in specimens under shear-dominated loading compared to tension-dominated loading before catastrophic failure;

6. the S-N curves clearly shows the detrimental effects of the shear load component on the fatigue life of all the investigated notched laminates since the slopes of S-N curves decrease with increasing the shear load component. On the other hand, the conditions to achieve  the endurance limit are approximately the same (40\% of $P_{max}$) for both notched and notch-free laminates at different multiaxiality ratios except for notched cross-ply laminates under tensile fatigue loading;

7. in this case, the endurance limit is about 55\% of $P_{max}$ which is much higher than the one of notch-free laminates featuring the same layup. This is probably due to the reduced stress concentration of a central crack caused by splitting occurring at the notch tip during the fatigue since the development of splitting is the main damage mechanism for this loading case;

8. the foregoing results are of utmost importance for the structural design of polymer matrix composites under multi-axial loading condition but so far rarely investigated. The lack of experimental data on this topic in the literature hindered the development of more accurate models which can guarantee a safe design in particular when composite structural components subject to multi-axial fatigue loading. As a first step towards filling the forgoing knowledge gap, this study provides a comprehensive experimental data on this topic.




\section*{Acknowledgments}
Marco Salviato acknowledges the financial support from the Haythornthwaite Foundation through the ASME Haythornthwaite Young Investigator Award and from the University of Washington Royalty Research Fund. This work was also partially supported by the Joint Center For Aerospace Technology Innovation through the grant titled ``Design and Development of Non-Conventional, Damage Tolerant, and Recyclable Structures Based on Discontinuous Fiber Composites".

\clearpage

\section*{Figures and Tables}

\newpage

\newpage

\newpage

\newpage

\newpage

\newpage

\newpage

\newpage

\newpage

\newpage

\newpage
\begin{table}[H]
\center
\scalebox{0.68}{
\begin{tabular}{ccc}
\hline
Stacking Sequences & Dimension [mm] & Measured Properties\\ \hline
$[0]_{8}$  & 190x20x1.7 & $E_{1}$=42.9 GPa, $f_{1t}$=900 MPa\\ 
$[90]_{8}$  & 210x30x1.7 & $E_{2}$=8.3 GPa, $f_{2t}$=43 MPa\\
$[+45/$-$45]_{2s}$ & 210x30x1.7 & $G_{12}$=5.8 GPa\\
$[0/90]_{2s}$ & 200x20x1.7 & $f_{t}$=540 MPa\\
$[+45/90/$-$45/0]_{s}$ & 200x20x1.7 & $f_{t}$=423 MPa\\ \hline
\end{tabular}}
\caption{Mechanical properties of unnotched specimens under tensile quasi-static loading condition. Note for the subscripts: 1 - axial direction, 2 - transverse direction.}
\label{tab:mechanicalproperties}
\end{table}

\begin{table}[H]
\center
\scalebox{0.58}{
\begin{tabular}{cccc}
\hline
Stacking Sequences & Multiaxiality Ratio & Normal Strength, $\sigma_{N,max}$ [MPa] & Shear Strength, $\tau_{N,max}$ [MPa]  \\ \hline
\multirow{5}{8em}{$[0/90]_{2s}$ with central crack} & 0 & 275.57 & 0 \\
& 0.262 & 213.02 & 57.08 \\
& 0.785 & 97.05 & 97.05 \\
& 1.309 & 25.96 & 96.87 \\
& 1.571 & 0 & 93.18 \\
\multirow{5}{8em}{$[+45/90/$-$45/0]_{s}$ with central crack} & 0 & 317.52 & 0 \\
& 0.262 & 289.99 & 77.70 \\
& 0.785 & 131.71 & 131.71 \\
& 1.309 & 36.57 & 136.48 \\
& 1.571 & 0 & 128.56 \\
\multirow{5}{8em}{$[+45/90/$-$45/0]_{s}$ with open hole} & 0 & 256.47 & 0 \\
& 0.262 & 217.24 & 58.21 \\
& 0.785 & 116.33 & 116.33 \\
& 1.309 & 34.76 & 129.72 \\
& 1.571 & 0 & 124.46 \\ \hline
\end{tabular}}
\caption{Average nominal normal and shear strengths of notched specimens under multi-axial quasi-static loading condition.}
\label{tab:notchedproperties}
\end{table}

\begin{table}[H]
\center
\scalebox{0.58}{
\begin{tabular}{ccccc}
\hline
Multiaxiality Ratio & Percentage of $P_{max}$ & $[0/90]_{2s}$ (crack) & $[+45/90/$-$45/0]_{s}$ (crack) & $[+45/90/$-$45/0]_{s}$ (hole)\\ \hline
\multirow{3}{1em}{0} & 70 \% & 1.8 $\times$ $10^5$ & 2.4 $\times$ $10^4$ & 5.2 $\times$ $10^4$ \\
& 55 \% & 2 $\times$ $10^6$ & 3.1 $\times$ $10^5$ & 2.9 $\times$ $10^5$ \\
& 40 \% & 2 $\times$ $10^6$ & 2 $\times$ $10^6$ & 2 $\times$ $10^6$ \\
\multirow{3}{1em}{0.262} & 70 \% & 5.8 $\times$ $10^4$ & 6617 & 5.4 $\times$ $10^4$ \\
& 55 \% & 4.1 $\times$ $10^5$ & 3.7 $\times$ $10^4$ & 1.4 $\times$ $10^5$ \\
& 40 \% & 2 $\times$ $10^6$ & 2 $\times$ $10^6$ & 2 $\times$ $10^6$ \\
\multirow{3}{1em}{0.785} & 70 \% & 4133 & 3117 & 2.8 $\times$ $10^4$ \\
& 55 \% & 7.7 $\times$ $10^4$ & 2.7 $\times$ $10^4$ & 7 $\times$ $10^4$ \\
& 40 \% & 1.4 $\times$ $10^6$ & 2 $\times$ $10^6$ & 2 $\times$ $10^6$ \\
\multirow{3}{1em}{1.309} & 70 \% & 5367 & 2741 & 1.3 $\times$ $10^4$ \\
& 55 \% & 4.8 $\times$ $10^4$ & 3.9 $\times$ $10^4$ & 1.2 $\times$ $10^5$ \\
& 40 \% & 2 $\times$ $10^6$ & 2 $\times$ $10^6$ & 2 $\times$ $10^6$ \\
\multirow{3}{1em}{1.571} & 70 \% & 1664 & 2493 & 3545 \\
& 55 \% & 7070 & 3.8 $\times$ $10^4$ & 5.5 $\times$ $10^4$ \\
& 40 \% & 2 $\times$ $10^6$ & 2 $\times$ $10^6$ & 2 $\times$ $10^6$ \\ \hline
\end{tabular}}

\caption{Average fatigue lifetimes measured from the multi-axial fatigue tests on both notched cross-ply and quasi-isotropic laminates. Note that two million cycles represents run-out.}
\label{tab:fatiguetimelifes}
\end{table}
\end{document}